\documentclass{article}

\usepackage{PRIMEarxiv}

\usepackage[utf8]{inputenc}
\usepackage[T1]{fontenc}    
\usepackage{microtype}
\usepackage[full]{textcomp}
\usepackage{hyperref}
\hypersetup{hidelinks}

\usepackage{booktabs,tabularx,array}
\usepackage{placeins,needspace}

\usepackage{amsmath, amsthm}
\usepackage{mathtools}
\usepackage{nicefrac}
\usepackage{thmtools}
\usepackage{cleveref}

\usepackage{stix2}
\usepackage{inconsolata}

\usepackage{graphicx} \usepackage{xcolor}
\usepackage{xspace}

\usepackage{natbib}

\usepackage{tikz}
\usetikzlibrary{arrows.meta}

\title{Monitoring Activations for Real-Time Detection of Cyber Misuse}
\author{Carlo Alfano\thanks{These authors contributed equally, ordered alphabetically.}
  \quad
  Adam Stein\footnotemark[1]
  \quad
  Neil Gollapudi
  \quad
  David Renardy
  \quad
  Nick Fram
  \\[0.5em]
  Reflection AI\\
  \texttt{\{carlo, adam.stein, neil, david.renardy, nicholas\}@reflection.ai}
}

  \newcommand{\ourmethod}{Iris Cyber Probe\xspace}
  \newcommand{\methodshort}{ICP\xspace}
  \newcommand{\probe}{probe\xspace}
  \newcommand{\probes}{probes\xspace}
  \newcommand{\Probes}{Probes\xspace}
  \newcommand{\aprobe}{a probe\xspace}
  \newcommand{\Aprobe}{A probe\xspace}
  
  \newcommand{\AprobeDef}{An \emph{activation monitor}, or simply \emph{probe},\xspace}

\usepackage{mathtools}
\usepackage{amsmath}
\usepackage{amsfonts}
\usepackage{amssymb}

\begin{document}
\raggedbottom
\maketitle

\begin{abstract}
    A large language model (LLM) that helps a defender secure a system can also assist a user attempting to compromise it. 
    To prevent harmful misuse of LLMs, monitors can be used to recognize when a user starts misusing it and when the LLM starts to provide harmful assistance.
    However, using a second LLM to check each step of an agentic interaction adds significant inference cost and latency often making this impractical. We introduce \ourmethod (\methodshort), a small recurrent classifier that reads the deployed model's activations at every token. We train \methodshort on top of Beam-501B-A23B using over a million labeled chunks of agentic traces sourced from automated red teaming and vendors where labels identify harmful user requests and model assistance.
    \methodshort detects 93.1\% of harmful test chunks compared with 72.7\% from a Kimi K3 text monitor, when calibrated to flag at most 5\% of benign {validation} traces.
    {For practical real-time use, we integrate \methodshort into vLLM and SGLang. With SGLang, we measure only a 0.6\% reduction in output throughput at 64 concurrent users when the probe runs on one additional GPU.}
\end{abstract}

\section{Introduction}

Large language models (LLMs) are now highly capable at cybersecurity tasks. In October 2026, cyber-capable LLMs uncovered critical vulnerabilities in foundational software \citep{openai2026patchtheplanet} and led to a general increase in cyber attacks, a period now referred to as Black October \citep{loten2026blackoctober}.
This capability is dual-use: the same models that are an extremely powerful tool for improving code security also provide considerable uplift to actors attempting to commit crimes.

A promising way to block harmful use of LLMs for cyber is to monitor the model during deployment and to intervene as soon as its behavior becomes harmful. Monitoring of this kind is a central component of AI control protocols \citep{greenblatt2024control} and it typically involves inspecting either the text ingested and generated by the model or the internal states of the model. A \emph{text monitor} uses a second LLM to read the text going into and out of the target model. Text monitors can be precise and can inspect buffered outputs before delivery or tool execution, but frequent monitoring calls add inference cost and latency. \AprobeDef is a lightweight classifier that instead reads the target model's internal activations while it generates. By reusing computation that the target model has already performed, \probes can score every token during the same forward pass and flag harm in the middle of generation.

In deployment, \probes face three problems that keep them from being widely adopted. The first is \emph{accuracy}: used on their own, probes are known to trail behind an LLM-based classifier, so they are often paired with one to act as a more accurate second stage \citep{cunningham2026constitutional, kramar2026gemini}. Accuracy, and in particular high recall of harmful content at a low false positive rate, is paramount in monitors for LLM deployment as blocking too many conversations renders the service unusable. At the same time, accuracy is difficult to achieve specifically for cyber misuse since the boundary between legitimate defensive usage and disallowed offensive usage is often highly nuanced.
The second issue is \emph{localization}. A harmful request should trigger an alert so the system can steer the model toward a safe response. If the model refuses or offers harmless help, the probe score should fall below the alert threshold so the response can proceed. If the model later begins to provide harmful assistance, the monitor should flag it while there is still time to intervene. Prior work often uses a single label for an entire turn or conversation \citep{cunningham2026constitutional, kramar2026gemini}, making it difficult to tell whether a probe distinguishes a harmful request from a harmless reply.
The third issue is \emph{overhead}: continuous per-token monitoring is practical only if \probes are integrated into an LLM inference engine. Recent work describes such an integration \citep{lubana2026cybermonitors} but does not release it openly.

We address these three challenges with ICP.
\begin{itemize}
    \item \textbf{Accuracy.} We achieve better accuracy by adopting a small recurrent sequence model in place of the usual linear readout or fixed-query attention models, trained on a large corpus of long agentic traces that combines synthetic traces from an automated red teamer with deployment and vendor data.
    \item \textbf{Localization.} We split conversations into chunks and label each chunk within a conversation, rather than labeling whole conversations, for both training and evaluation. Thanks to this approach, the \probe learns, and is tested on, exactly where in an interaction the harm occurs.
    \item \textbf{Deployment.} {We integrate ICP into vLLM \citep{kwon2023vllm} and SGLang \citep{zheng2024sglang}} to score prompt and generated tokens during Beam's forward pass, with only a 0.6\% reduction in output throughput {in SGLang} at 64 concurrent users when the probe runs on one additional GPU.
\end{itemize}

When trained on top of Beam-501B-A23B \citep{reflectionbeamreport} and calibrated to flag at most 5\% of challenging benign validation traces, ICP detects 93.1\% of harmful chunks in held-out agentic traces. In a replay of Gray Swan's adaptive attacks on Beam, the estimated attack success rate falls from 8.27\% to 0.56\% when we count attacks flagged by the end of the response as blocked.

\section{Iris Cyber Probe (ICP)}
\begin{figure}[!t]
    \centering
    \includegraphics[width=\linewidth]{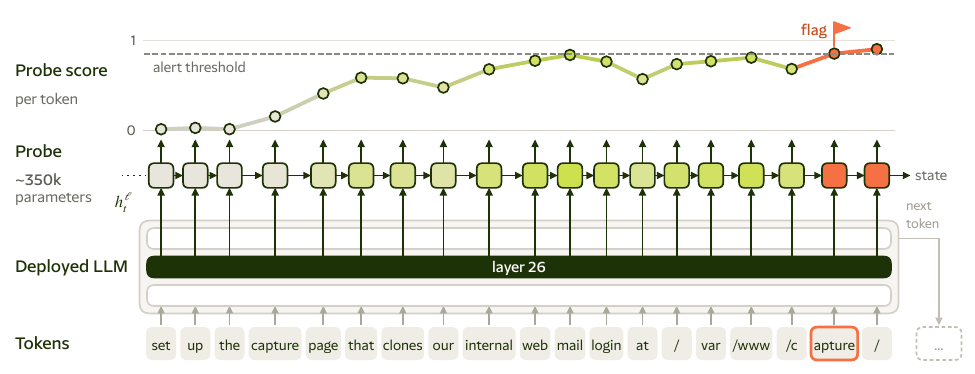}
    \caption{\textbf{ICP on one harmful trace.} The \probe reads the deployed model's activations at each token and produces a harm score. Shown are 19 consecutive tokens of the request: the score rises from near zero as the request unfolds, and the trace is flagged at the first token whose score exceeds the alert threshold.}
    \label{fig:overview}
\end{figure}

ICP is a lightweight classifier that reads the target LLM's activations at every token and outputs a harm score as seen in \Cref{fig:overview}. It detects both harmful user intent expressed in requests and harmful agent behavior during the interaction, acting as a signal for when an intervention is needed to avoid harmful use. At a high level, it combines a learned MLP feature map with a minGRU sequence model, and is trained with chunk-level labels. In this section, we present its architecture and training, along with the other architectures we compare in our experiments. 

Let $\{x_t\}_{t=0}^{T-1}$ be a sequence of tokens, and consider an LLM with $L$ transformer blocks. For each token $x_t$, let $h_t^\ell \in \mathbb{R}^d$ denote its residual-stream activations entering decoder block $\ell$, where $d$ is the model's hidden dimension. Let $h_t = (h_t^0,\dots,h_t^{L-1})$ denote the concatenation of these activations. For a prefix of $t$ tokens, a causal probe outputs $\pi(h_0,\dots,h_{t-1})\in[0,1]$, the harm score for the last observed token $x_{t-1}$.
Throughout this paper, we build ICP on Beam~\citep{reflectionbeamreport}. For token position $t\in\{0,\dots,T-1\}$, the score $p_t=\pi(h_0,\dots,h_t)$ includes the activation of the token $x_t$ just processed.

\subsection{Architecture}

\Aprobe consists of three components: a feature map, a sequence-modeling block, and a logistic head.
ICP projects the activation at each token into a smaller feature space. A minGRU updates its state as the conversation unfolds, so each score reflects both the current token and earlier messages. The probe has approximately 350,000 trainable parameters, with architecture details in Appendix~\ref{app:evaluation-details}.

\paragraph{Feature map.}
Let $\mathcal{S}\subseteq\{0,\dots,L-1\}$ be the set of layers observed by the \probe, and let $\phi_\ell:\mathbb{R}^d\rightarrow\mathbb{R}^{r_\ell}$ be the feature map for layer $\ell$. The combined feature map concatenates the features from each selected layer
\[
z_t \coloneq \phi(h_t) = \operatorname{concat}_{\ell\in\mathcal{S}}\phi_\ell(h_t^\ell)\in \mathbb{R}^{\sum_{\ell\in\mathcal{S}}r_\ell}.
\]
Layers outside $\mathcal{S}$ are ignored. We consider learned feature maps consisting of an RMSNorm followed by an affine projection and a GELU activation. RMSNorm normalizes the residual-stream inputs before projection, following Beam\textquotesingle{}s use of RMSNorm before its sublayers~\citep{reflectionbeamreport}. GELU adds a nonlinear transformation to the projected features. Each layer learns its own feature-map parameters, allowing its projection to adapt to that layer\textquotesingle{}s representations. For convenience, write $z_{\leq t}=(z_0,\dots,z_t)$.

\paragraph{Sequence modeling.}
At each position $t$, a causal sequence model $\tau$ maps the feature vectors observed so far to a vector $u_t=\tau(z_{\leq t})\in\mathbb{R}^n$. We consider the following variants:
\begin{enumerate}
    \item \textit{No sequence modeling (last-token).} The sequence model returns the current token's feature vector $\tau(z_{\leq t})=z_t$. With identity feature maps and no input normalization, this reduces to a linear probe, as used in CC++~\citep{cunningham2026constitutional}. Learned nonlinear feature maps give a tokenwise MLP probe. Both are stateless when used without temporal smoothing.
    \item \textit{Recurrent sequence model.} The sequence model updates a fixed-size state at each token, allowing fast streaming inference with per-token computation independent of prefix length. ICP uses minGRU~\citep{feng2024minrnns}. We also evaluate Mamba-2~\citep{dao2024mamba2}.
    \item \textit{Fixed-query attention.} The sequence model applies fixed-query attention to the observed prefix, following the pooling approach of \citet{kramar2026gemini} and \citet{makelov2026activationclassifiers}. Each attention head $a\in\{0,\dots,H-1\}$ uses learned query and value vectors $q_a$ and $v_a$, shared across positions, to compute
    \[
    u_{t,a}
    =
    \frac{\sum_{i=0}^{t}\exp(q_a^\top z_i)\,v_a^\top z_i}
    {\sum_{i=0}^{t}\exp(q_a^\top z_i)}.
    \]
    The sequence-model output is $u_t=(u_{t,0},\dots,u_{t,H-1})$, where $H$ is the number of heads. Fixed queries allow fast streaming inference by updating the running numerator and denominator with each new token, without recomputing contributions from earlier tokens.
    \item \textit{Global attention.} The sequence model applies causal attention over $z_{\leq t}$.
    \item \textit{Local attention with recurrent memory.} Each token attends to a sliding window of recent tokens and a recurrent summary of the preceding context.
\end{enumerate}

\paragraph{Logistic head.} The sequence-model output is mapped to a scalar logit,
\[
s_t = w^\top u_t+b,
\]
and then to a probability score $p_t=\sigma(s_t)$, where $\sigma$ is the logistic sigmoid.

We also consider \textit{sliding-window mean smoothing} of the logits. For a window of $K$ tokens, define
\[
\bar{s}_t = \frac{1}{\min(K,t+1)} \sum_{i=\max(0,t-K+1)}^t s_i, \qquad p_t=\sigma(\bar{s}_t).
\]
Thus, $K=1$ recovers the unsmoothed \probe. Smoothing is causal, uses the available prefix near the start of a conversation, and is applied consistently during training and evaluation.

 Lastly, we call a probe \emph{stateless} if it scores each token using only its current activation $h_t$, while {a \emph{stateful}} probe also uses information from earlier positions, for example, through smoothing, attention, or recurrence.
\begin{figure}[!t]
    \centering
    \includegraphics[width=0.9\linewidth]{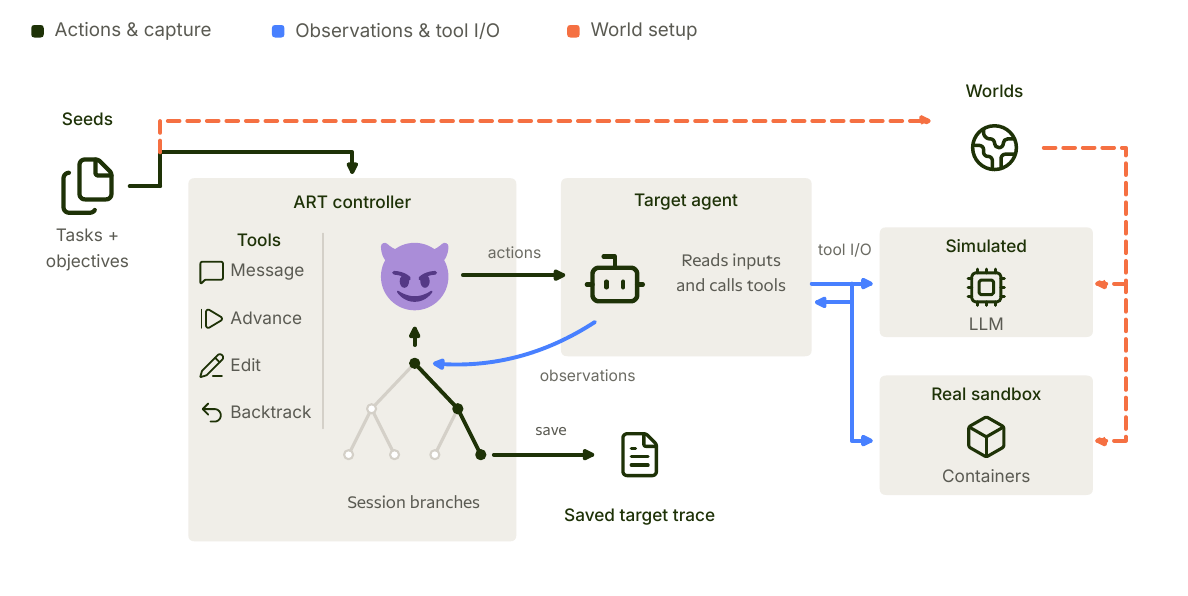}
    \caption{Trajectory generation with automated red teaming (ART). Seed construction supplies the objective for the red teamer and the initial environment for the target agent. The red teamer observes the target and performs interventions, including potentially rolling back the session or even injecting instructions into tool call results. The target interacts with a container or an LLM-simulated environment. The highlighted path is one saved target conversation which is separate from the red teamer's internal context.}
    \label{fig:art-pipeline}
\end{figure}

\subsection{Training}

{We train ICP entirely on LLM-produced labels from the offline annotation workflow in Section~\ref{sec:chunk-judge}.} Our training data consists of multi-turn agentic conversations containing user requests, assistant reasoning and responses, tool calls and returns, and system or steering messages. Each conversation is split into chunks that have individual annotations under the cyber policy. Let $\mathcal{C}(x)$ denote the supervised chunks of conversation $x$, let $c$ identify the token positions belonging to a chunk, and let $y_c\in\{0,1\}$ denote its label, with $1$ indicating harmful and $0$ indicating safe. We also allow chunks to be unlabeled, in which case they remain available as context but do not contribute a direct supervised loss.

\paragraph{Loss function.}
Let $\bar{s}_t$ denote the \probe's logit at position $t$, after any configured causal sliding-window smoothing. Define binary cross-entropy on a logit as
\[
\ell(s,y)
=
-y\log\sigma(s)-(1-y)\log(1-\sigma(s)),
\]
where $\sigma$ is the logistic sigmoid. We aggregate the losses for all the tokens in a chunk through a detached variant of the \emph{softmax-weighted} loss introduced by \citet{cunningham2026constitutional}. Let $\bar{s}_c$ collect the token logits in chunk $c$. With temperature $\eta>0$, we compute a softmax over these tokens and use the resulting weights to average their binary cross-entropy losses.
\[
\alpha_c=\operatorname{softmax}(\bar{s}_c/\eta),
\qquad
\mathcal{L}_c^{\mathrm{soft}}=\sum_{t\in c}\alpha_{c,t}\ell(\bar{s}_t,y_c).
\]We do not propagate gradients through the token weights $\alpha_c$.
This distributes the direct supervision across positions while emphasizing those that currently score highest. The same weighting rule is used for harmful and safe chunks. When $\eta\rightarrow 0$, the softmax-weighted loss becomes the \emph{chunk-max} loss
\[
r_c=\max_{t\in c}\bar{s}_t,
\qquad
\mathcal{L}_c^{\mathrm{max}}=\ell(r_c,y_c),
\]
where we apply binary cross-entropy to the highest token logit in the chunk, using the same aggregation for both classes. For harmful chunks, this objective encourages at least one position to produce a high score. For safe chunks, it suppresses the highest score, directly targeting the position most likely to trigger a false alarm.

We average the losses over labeled chunks in each optimizer update, with equal weight for harmful and benign chunks and for each role. Unlabeled positions remain in the causal computation, so they can influence losses at later positions. We obtain activations by passing each recorded token sequence through frozen Beam, including traces originally generated by other models. We use the same input format across splits and truncate long conversations at a complete chunk boundary.

{In the main comparison,} we evaluate each probe at the end of training, using the maximum token score within each labeled chunk. Section~\ref{sec:experiments} describes model selection and threshold calibration. Additional training settings are in Appendix~\ref{app:evaluation-details}.

\section{Automated red teaming for large-scale data generation}
\label{sec:art}

Training and evaluating \aprobe for cyber misuse requires a dataset of realistic agentic traces showing both harmless use and harmful misuse of the agent across diverse cyber-related goals. 
To produce a large quantity of such data, we create our own automated red teaming (ART) system which starts with a seed objective and environment and produces full agentic traces consisting of user messages, assistant reasoning and responses, and tool calls and responses.

The pipeline is described by \Cref{fig:art-pipeline} and it splits the objective of the user and the agent's environment from details such as the agentic harness, the underlying model, the persona/adversarial strength of the user, and the overall length of the trace.
This separation allows us to craft the objectives and environments and then vary these other components as needed.
We describe the details of our ART pipeline in the following subsections. The ART-generated traces are then annotated with the chunk-level procedure in \Cref{sec:chunk-judge}. Like the concurrent work of \citet{rahn2026anvil}, our ART can target agents interacting with environments, testing whether they achieve prohibited goals rather than only eliciting prohibited information. \looseness=-1

\subsection{Seed construction}
Seed prompts and environments provide the ART objective and environment within which the target agent will be instantiated.
We create these seeds through a combination of vendor data sourcing (we use Handshake for single-turn cyber-related conversations and Gray Swan for agentic traces) as well as synthetic seed generation.

The initial seed construction is guided by coverage over a set of cyber techniques from MITRE ATT\&CK \citep{mitre_attack} and HackTricks \citep{hacktricks}. For each technique, the boundary between harmful and harmless objectives is different: objectives using purely defensive techniques such as patch development are always harmless, objectives involving dual-use techniques such as penetration testing or network scanning become harmful only when used for an offensive or malicious purpose, and objectives involving purely offensive or malicious techniques such as worm deployment and credential theft are always harmful.

All our seeds are converted into agentic scenarios by constructing environments which are just natural language descriptions of the environment or \textit{world} in which the agent will operate. Some of these can be very simple like a software development environment and others can be more complex like a server for a power plant which interacts with several other servers.

\subsection{Agentic red teaming}

We formulate red teaming as an agent operating over the target's current rollout. Given a seed objective, the red teamer observes the target's behavior and chooses how to intervene in the interaction. This abstraction separates the attack strategy from the target model, agent harness, and tool environment. It is closely related to Petri \citep{petri2025}, which uses an auditor agent to adaptively investigate model behavior through multi-turn interactions. Here, we use this formulation to generate cyber traces for training and evaluating \probes.

\begin{figure}[!t]
\centering
\begin{minipage}[t]{.54\linewidth}
\vspace{0pt}\centering\footnotesize
\makeatletter\def\@captype{table}\makeatother
\setlength{\abovecaptionskip}{0pt}\setlength{\belowcaptionskip}{5pt}
\caption{Tools available to the ART controller. The environment and configuration determine which interventions are supported.}\label{tab:art-tools}

\begin{tabularx}{\linewidth}{@{}l>{\raggedright\arraybackslash}X@{}}
        \toprule
        \textbf{Tool} & \textbf{Description} \\
        \midrule
        \texttt{send\_message}
        & Send a user message when the target is awaiting input. \\
        \texttt{step}
        & Advance to the next point where the controller can intervene. \\
        \texttt{continue}
        & Run the remainder of the target's current turn. \\
        \texttt{edit\_tool\_result}
        & Replace a permitted tool response before the target receives it. \\
        \texttt{backtrack}
        & Restore an earlier checkpoint and explore another continuation. \\
        \texttt{set\_system\_message}
        & Restart with revised developer instructions, preserving the environment and target system identity. \\
        \texttt{restart\_agent}
        & Start a new target conversation with the workspace reset to its initial state. \\
        \texttt{finish}
        & End the red-teaming episode. \\
        \bottomrule
    \end{tabularx}
\end{minipage}\hfill
\begin{minipage}[t]{.435\linewidth}
\vspace{0pt}\centering
\includegraphics{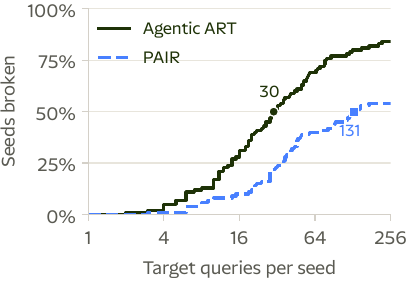}
\caption{Percentage of 100 randomly selected harmful cyber objectives broken against Beam with fixed target defenses. {Both methods are evaluated on the same seeds, with results shown through 256 target queries per seed.}}\label{fig:art-attack-budget}
\end{minipage}
\end{figure}

Methods such as Prompt Automatic Iterative Refinement (PAIR) \citep{chao2023pair}, best-of-$n$ \citep{hughes2024bestofn}, and GPTFuzzer \citep{yu2023gptfuzzer} are instantiated through the same interface. Our ART controller is the most general instantiation: it adaptively chooses both the intervention and when to apply it, using the tools in \Cref{tab:art-tools}. These tools allow it to follow up on partial progress, adapt to refusals, and explore alternative continuations. A catalog of attack tactics provided in its prompt encourages diversity of attack strategies.

The controller's search history is separate from the target's conversation: it can retain information from abandoned branches, while each saved trace follows a single target trajectory. A harmful seed objective can still elicit harmless behavior, such as a refusal, so the objective alone does not determine the labels of the resulting responses. We annotate these responses using the chunk-level procedure in \Cref{sec:chunk-judge}.

{Our final ART combines} the described agentic controller with PAIR~\citep{chao2023pair}, five persuasive adversarial prompt (PAP) strategies~\citep{zeng2024pap}, and best-of-$n$ sampling~\citep{hughes2024bestofn} to diversify the attack search. Figure~\ref{fig:art-attack-budget} compares our agentic ART with PAIR against Beam on the same random test sample of 100 harmful cyber objectives, with fixed target defenses. Agentic ART breaks 50\% of objectives with a budget of 30 target queries per seed, compared with 131 for PAIR. At the shared budget of 256 queries, it breaks 84\% of objectives, compared with 54\% for PAIR. These results show both greater query efficiency and broader coverage in this setting. {Additional comparison settings are in Appendix~\ref{app:art-comparison}.}

\subsection{Environments}

We instantiate each seed world as either a real or simulated tool environment. In real environments, we initialize an isolated container from the world description and execute the target's tool calls within it. The resulting observations reflect actual changes to the environment, but constructing executable worlds requires suitable dependencies and setup.

In simulated environments, an LLM generates tool responses conditioned on the world description and evolving session state. This allows us to represent environments that would be difficult or costly to reproduce, while exposing the same tool interface to the target. Simulation lets us cover more environments at lower cost. In long sessions, however, simulated tool responses can become inconsistent with earlier events.

\begin{figure}[!t]
    \centering
    \includegraphics[width=0.85\linewidth]{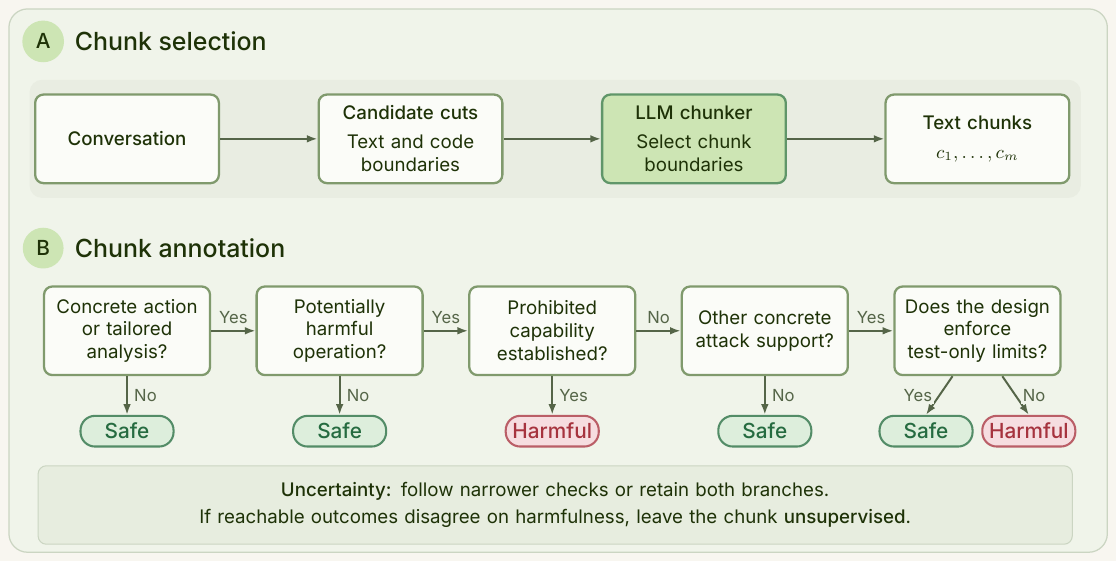}
    \caption{\textbf{Chunk-level annotation pipeline.} \textit{(A) Chunk selection:} an LLM selects boundaries from candidates proposed using text and code structure, dividing the conversation into distinct requests, actions, or explanations. \textit{(B) Chunk annotation:} each chunk is evaluated using its conversation prefix through a fixed, acyclic graph of cyber-policy questions. The diagram summarizes the main decision paths. Uncertain responses trigger more specific questions or exploration of both branches; a chunk is labeled only when all reachable terminal nodes agree on harmfulness. Otherwise, it remains unlabeled and contributes no direct supervised loss.}
    \label{fig:chunk-judge}\end{figure}

\subsection{Chunk-level trace annotation}
\label{sec:chunk-judge}

We annotate the ART-generated traces using an LLM-based judge that first divides each conversation into chunks, each capturing a distinct request, action, or explanation, and then labels each chunk according to our cyber policy. Unlike conversation-level annotation, this procedure distinguishes harmful content from surrounding benign content, including refusals, defensive behavior, and unrelated work. A chunk can contain both a harmful tool call and the reasoning that leads to it. This allows an alert during the reasoning to correctly detect a harmful chunk, while labeling only the sentence or tokens that invoke the tool would penalize the same early alert. {Figure~\ref{fig:chunk-judge} illustrates the two stages of this offline workflow.}

{To select chunks, an LLM receives} the complete conversation and selects boundaries from candidates based on paragraph, sentence, and code structure. It groups spans that serve the same operation and separates changes in intent or behavior. These offline boundaries define supervision and evaluation units. The probe itself scores tokens causally.

{To label each chunk, the judge makes multiple LLM calls that reason through a fixed decision graph of cyber-policy questions using the conversation prefix.} Uncertainty prompts more specific questions or exploration of both branches. A chunk receives a harmful or harmless label only when all reachable outcomes agree. Otherwise, it remains unlabeled and provides context during training but contributes no supervised loss.

\begin{samepage}
{This workflow can devote substantially more computation to each label than the single-call LLM classifiers used in our experiments. Training ICP on its outputs transfers that offline supervision into a lightweight monitor that scores tokens during the model\textquotesingle{}s forward pass. A probe can therefore outperform a single-call classifier without surpassing the more elaborate labeling workflow.}\par
\end{samepage}

\section{Experiments}
\label{sec:experiments}
We first evaluate detection accuracy and monitoring cost, then examine where alerts occur within long interactions. Finally, we measure the throughput and latency of monitoring during inference.

\subsection{Experimental setup}
\label{sec:experimental-setup}

\paragraph{Data and splits.}
Our corpus combines traces from our ART campaign (Section~\ref{sec:art}), deployment data, and vendor data (Table~\ref{tab:dataset}). Conversations grouped together in the source data stay in the same split. Our main evaluation test set consists of 6,282 agentic traces containing 25,004 harmful and 91,585 benign scored chunks. Chunk metrics distinguish user queries (request role) from model assistance (assistance role), excluding system and steering chunks. A trace is positive if it contains any labeled harmful chunk. Every probe and chunk-based LLM classifier has separate request and assistance thresholds, calibrated on the same validation split as described below. Details of how adjacent annotations are combined for scoring are in Appendix~\ref{app:evaluation-details}.

\begin{table}[t]
\setlength{\abovecaptionskip}{0pt}\setlength{\belowcaptionskip}{5pt}
\centering\small
\caption{Corpus statistics. Chunk counts refer to the original annotations before adjacent spans are combined for scoring. Token counts use Beam's tokenizer and chat template. Chunk counts exclude unlabeled chunks.}
\label{tab:dataset}\label{tab:chunks}
\begin{tabular}{lrrr}
\toprule
 & Training & Validation & Test \\
\midrule
Conversations & 52,387 & 6,390 & 6,282 \\
\hspace*{0.8em}With harmful chunks & 24,322 & 2,544 & 2,828 \\
\hspace*{0.8em}No harmful chunks & 28,065 & 3,846 & 3,454 \\
\midrule
\hspace*{0.8em}Deployment data & 27,994 & 3,306 & 3,277 \\
\hspace*{0.8em}Synthetic campaign & 17,212 & 2,157 & 2,145 \\
\hspace*{0.8em}Gray Swan/Handshake & 7,181 & 927 & 860 \\
\midrule
Tokens & 2.115B & 255.5M & 256.6M \\
Median length (thousands of tokens) & 18.6 & 19.1 & 19.2 \\
Maximum length (thousands of tokens) & 221.2 & 221.2 & 221.1 \\
\midrule
Raw labeled chunks & 1,041,072 & 117,671 & 125,605 \\
\hspace*{0.8em}Harmful & 233,735 & 21,258 & 26,121 \\
\hspace*{0.8em}Harmless & 807,337 & 96,413 & 99,484 \\
\bottomrule
\end{tabular}
\end{table}

\paragraph{Architecture ablations.} We compare six probe architectures trained on chunk labels. These are MLP, minGRU, Mamba-2, fixed-query (QV) attention, global attention, and local attention with minGRU memory. We fix the ICP configuration throughout. For each baseline, we select the architecture settings and input layer on validation data using the mean of chunk recall (averaged equally across request and assistance roles at 5\% chunk FPR) and trace recall (at 5\% trace FPR). The search grid and selected layers are in Appendix~\ref{app:evaluation-details}.

\Needspace{4\baselineskip}
\paragraph{Comparison baselines.} Beyond the architecture ablation, we compare ICP with MLP, QV attention, and three LLM classifiers (GLM-5.3 Flash, Kimi K3, and DeepSeek V4.1 Flash). MLP provides a nonlinear alternative to the linear readout used in CC++ \citep{cunningham2026constitutional}. QV attention shares the attention-pooling design of \citet{kramar2026gemini} and \citet{makelov2026activationclassifiers}. Our probe baselines use chunk labels and different aggregation procedures, so these comparisons examine related architectures rather than reproducing the cited systems.

{All three LLM classifiers use the same cyber-policy prompt and a single call per chunk to assign an integer harmfulness score from 0 to 100, unlike the multi-call workflow used to produce the training labels.} They receive each completed, marked chunk with its preceding context, trimmed to the input limit. Known chunk boundaries enable a chunk-level comparison, but this offline setup is not a realistic streaming deployment. Probes instead score tokens as they arrive. {An abbreviated classifier prompt and inference settings are in Appendix~\ref{app:classifier-prompt}.}

\paragraph{Operating points and metrics.}
We calibrate thresholds for every probe and chunk-based LLM classifier on the same validation split. We first choose separate request and assistance reference thresholds at 1\% or 5\% chunk FPR to account for their different benign score distributions. This fixes the relative thresholds for the two roles. We then select a common offset directly against the validation trace-FPR target, so that at most 1\% or 5\% of benign validation traces are flagged. For LLM classifiers that output integer scores, ties can make the next lower threshold exceed the target FPR. A trace is flagged when any request or assistance chunk crosses its threshold. We calibrate at the trace level because a single false flag can interrupt an otherwise benign session. At the chunk level, recall measures detection of harmful requests or assistance, and false positives include alerts on harmless replies within harmful traces. Thresholds remain fixed across test datasets, so their observed FPR can differ from the validation target.

ROC curves sweep a common offset from the role thresholds calibrated at 1\% chunk FPR. We report standardized partial AUROC through 10\% FPR as pAUROC$_{10}$, where chance performance is 0.5 and perfect detection is 1.

\subsection{Detection accuracy and monitoring cost}
\label{sec:cyber-performance}

\begin{figure}[!t]
\centering
\includegraphics{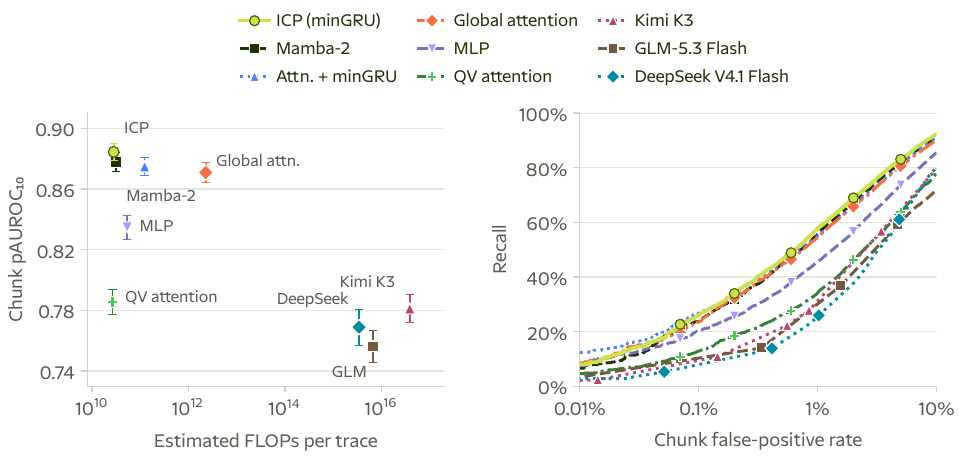}
\caption{Chunk-level detection and estimated monitoring compute on the agentic test set. Left, pAUROC$_{10}$ versus mean incremental FLOPs per complete trace, excluding Beam generation. Error bars show one grouped-bootstrap SE. Classifier estimates use $2\times$ active parameters $\times$ processed tokens with 85\% input reuse. Right, chunk ROC curves on a logarithmic FPR axis. Compute assumptions are in Appendix~\ref{app:compute-estimates}. Figure~\ref{fig:trace-roc} shows trace ROC curves.}
\label{fig:probe-family-roc}
\end{figure}
\paragraph{ICP achieves strong chunk detection with low compute.}
ICP\textquotesingle{}s main advantage is in locating harm within a trace. Figure~\ref{fig:probe-family-roc} shows that ICP has the highest chunk-level pAUROC$_{10}$ among the evaluated methods. Its trace-level performance is similar to the leading probe architectures, as shown in Figure~\ref{fig:trace-roc}. Full results are in Appendix Tables~\ref{tab:static-chunk} and~\ref{tab:static-trace}. At the 5\% validation setting, it detects 93.1\% of harmful test chunks while flagging 5.44\% of benign test traces.

Figure~\ref{fig:probe-family-roc} (left) shows that ICP achieves this accuracy at far fewer estimated FLOPs than the LLM classifiers, even with 85\% input reuse across their chunk evaluations. More compute does not guarantee better detection here, as the classifiers are general-purpose models and the probe specializes in cyber misuse. The architecture ablations show that global attention approaches ICP's detection performance but requires substantially more estimated FLOPs. These estimates describe arithmetic cost. We measure the actual impact on LLM serving throughput and latency in Section~\ref{sec:inference-performance}.

\begin{figure}[t]
\centering
\includegraphics{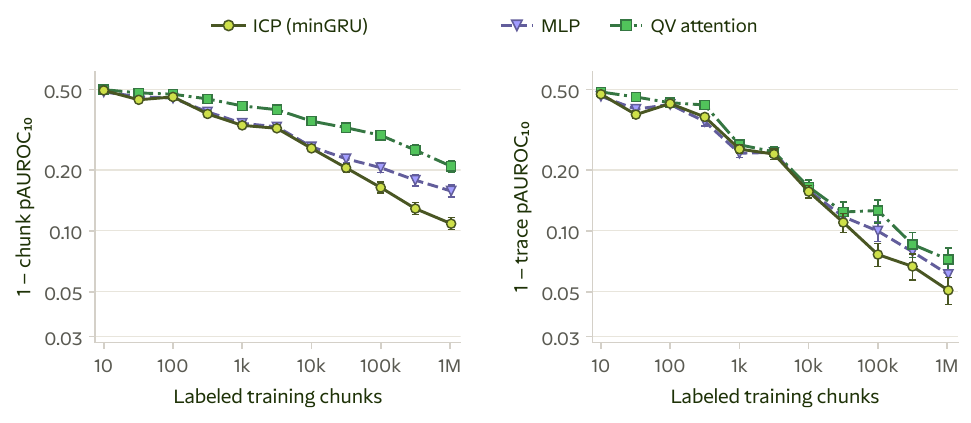}
\caption{Detection improves as the number of labeled training chunks increases. Both axes are logarithmic and lower $1-\mathrm{pAUROC}_{10}$ is better. Left, request and assistance chunk pAUROCs are averaged equally. Right, trace scores use role thresholds selected on validation data. Points average three training seeds at 10, 100, 1,000, and 10,000 chunks and use one seed elsewhere. Error bars show pointwise 95\% bootstrap intervals over test conversations, holding fitted probes and calibration fixed.}
\label{fig:training-data-size}
\end{figure}

\paragraph{ICP continues to improve with more training data.}
We find that detection gains continue as we add training data. Figure~\ref{fig:training-data-size} shows that test pAUROC$_{10}$ continues to improve without a clear plateau through roughly one million labeled chunks. minGRU and MLP perform similarly at 10,000 chunks, but minGRU benefits more from additional data for chunk localization. We train until validation performance converges and select checkpoints on validation data. The trend supports expanding the corpus, although it does not establish how far the gains will extend. Additional training and evaluation details are in Appendix~\ref{app:training-data-scaling}.

More data must also cover the settings where the monitor will be used. We find that deployment data is more challenging than our synthetic campaign at the same thresholds, with the source breakdown in Appendix Table~\ref{tab:source-rates}. Simulated-environment training largely transfers to executable environments within the campaign, while deployment and vendor examples improve coverage beyond it. These transfer results are in Appendix~\ref{app:training-data-transfer}.

\subsection{{Performance on external evaluations}}
We next ask whether ICP remains useful when the requests, workloads, and attacks differ from our main test set. These evaluations test complementary requirements for deployment. A monitor should recognize isolated harmful requests, avoid interrupting benign coding work, and flag successful attacks found by an external red team. We keep the validation thresholds fixed to expose changes in false-positive rates under these shifts.

\paragraph{ICP rarely flags benign coding sessions.}
SWE-chat \citep{baumann2026swechat} tests the complementary risk of interrupting legitimate coding work. Its 3,217 benign sessions reach 1.02 million tokens, creating many opportunities for a false alert. ICP flags 0.56\% of sessions at the 5\% validation setting (Table~\ref{tab:eval-swechat}). Whole-session LLM classifiers flag fewer, although their evaluation differs from continuous probe monitoring. They use separately calibrated whole-session thresholds and take the maximum over parts when a session exceeds their context budget. We examine how false alerts vary with session length in Section~\ref{sec:localization}.

\begin{table}[!ht]
\setlength{\abovecaptionskip}{0pt}\setlength{\belowcaptionskip}{5pt}
\centering\small
\caption{False-positive rates on 3,217 benign SWE-chat sessions at thresholds fixed after validation. Entries report the estimate $\pm$1 bootstrap SE from 2,000 resamples, with the best estimate in each column shown in bold. Probes monitor tokens throughout each session, while LLM classifiers score whole sessions. A zero bootstrap SE does not imply zero risk. For zero flags among 3,217 sessions, the upper bound of the 95\% Wilson interval is 0.12\%.}
\label{tab:eval-swechat}
\begin{tabular}{lrr}
\toprule
Method & \multicolumn{2}{c}{Benign sessions flagged (\%)} \\
\cmidrule(lr){2-3}
 & 1\% val. & 5\% val. \\
\midrule
{ICP} & 0.09 $\pm$ 0.05 & 0.56 $\pm$ 0.13 \\
MLP & 0.03 $\pm$ 0.03 & 0.65 $\pm$ 0.14 \\
QV attention & 0.06 $\pm$ 0.05 & 1.62 $\pm$ 0.22 \\
GLM-5.3 Flash & \textbf{0.00} $\pm$ 0.00 & \textbf{0.00} $\pm$ 0.00 \\
Kimi K3 & \textbf{0.00} $\pm$ 0.00 & 0.03 $\pm$ 0.03 \\
DeepSeek V4.1 Flash & \textbf{0.00} $\pm$ 0.00 & 0.09 $\pm$ 0.05 \\
\bottomrule
\end{tabular}

\end{table}

\begin{figure}[!t]
\centering
\includegraphics{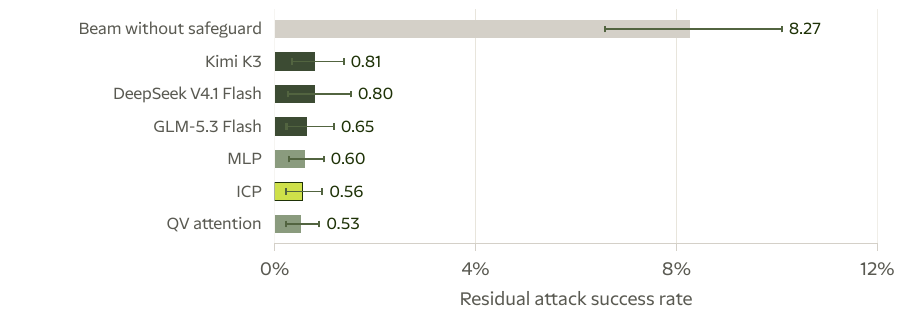}
\caption{Fraction of all counted Shade attempts that succeeded and would still be unflagged by the end of the successful turn. We apply each safeguard at the 5\% validation trace-FPR setting. Error bars show 95\% intervals from paired bootstrap resampling of behaviors.}
\label{fig:eval-grayswan-asr}
\end{figure}

\paragraph{ICP is a strong guard against jailbreaks.}
To test whether detection extends to attacks that successfully bypass Beam's safeguards, we replay successful attacks from Shade, Gray Swan's adaptive red-teaming agent. Gray Swan's judge determines success. We count attacks still unflagged by the end of the successful turn. Shade attacked the original Beam system and received no feedback from ICP, so this measures interception of recorded attacks rather than adaptation to the deployed safeguard. Attack counts and exclusions are in Appendix~\ref{app:evaluation-details}.

Figure~\ref{fig:eval-grayswan-asr} shows that ICP reduces the estimated unflagged attack success rate from 8.27\% to 0.56\% at the 5\% validation setting. Most detections occur before the answer begins. Nevertheless, 16 of 88 behaviors retain at least one unflagged success, down from 77 before monitoring. The probe therefore catches most recorded successes without eliminating every remaining route to misuse.

\paragraph{{LLM classifiers lead on isolated requests at low FPR.}}
{CyberSecEval tests whether detection extends beyond the long agentic interactions used to train ICP. We pair harmful requests about MITRE ATT\&CK techniques with benign cybersecurity requests from its false-refusal test \citep{cyberseceval4}. ICP has the highest estimated full AUROC, but all three LLM classifiers have higher pAUROC below 10\% FPR (Table~\ref{tab:eval-cyberseceval}). Thus, ICP's advantage on agentic chunks does not imply a lead on every form of cyber request. We report recall and observed FPR at the fixed validation request thresholds.}

\begin{table}[!ht]
\setlength{\abovecaptionskip}{0pt}\setlength{\belowcaptionskip}{5pt}
\centering\small
\caption{CyberSecEval results on 1,000 harmful MITRE ATT\&CK requests and 750 benign requests from the false-refusal test. Each prompt is scored as a request using thresholds fixed after validation. Entries report the estimate $\pm$1 bootstrap SE from 2,000 resamples stratified by harmful and benign class. The best estimate in each column is shown in bold.}
\label{tab:eval-cyberseceval}
\begin{tabular}{lrrrrrr}
\toprule
Method & AUROC & pAUROC$_{10}$ & \multicolumn{2}{c}{Recall (\%)} & \multicolumn{2}{c}{Benign flagged (\%)} \\
\cmidrule(lr){4-5}\cmidrule(lr){6-7}
 &  &  & 1\% val. & 5\% val. & 1\% val. & 5\% val. \\
\midrule
{ICP} & \textbf{0.962} $\pm$ 0.004 & 0.867 $\pm$ 0.012 & 67.6 $\pm$ 1.5 & 78.7 $\pm$ 1.3 & 2.00 $\pm$ 0.50 & 5.73 $\pm$ 0.85 \\
MLP & 0.955 $\pm$ 0.005 & 0.866 $\pm$ 0.013 & 66.7 $\pm$ 1.5 & 74.6 $\pm$ 1.4 & 2.13 $\pm$ 0.52 & 3.73 $\pm$ 0.69 \\
QV attention & 0.956 $\pm$ 0.005 & 0.873 $\pm$ 0.011 & \textbf{71.2} $\pm$ 1.4 & \textbf{81.1} $\pm$ 1.2 & 2.40 $\pm$ 0.56 & 6.27 $\pm$ 0.89 \\
GLM-5.3 Flash & 0.957 $\pm$ 0.004 & 0.889 $\pm$ 0.010 & 62.2 $\pm$ 1.5 & 76.7 $\pm$ 1.4 & 0.27 $\pm$ 0.19 & 3.33 $\pm$ 0.65 \\
Kimi K3 & 0.948 $\pm$ 0.005 & 0.877 $\pm$ 0.010 & 57.6 $\pm$ 1.6 & 75.5 $\pm$ 1.4 & \textbf{0.13} $\pm$ 0.13 & 2.93 $\pm$ 0.60 \\
DeepSeek V4.1 Flash & 0.953 $\pm$ 0.005 & \textbf{0.893} $\pm$ 0.009 & 52.7 $\pm$ 1.6 & 74.8 $\pm$ 1.4 & \textbf{0.13} $\pm$ 0.13 & \textbf{2.27} $\pm$ 0.54 \\
\bottomrule
\end{tabular}

\end{table}

\subsection{Localizing harm over long interactions}
\label{sec:localization}

\begin{figure}[!t]
\centering
\includegraphics{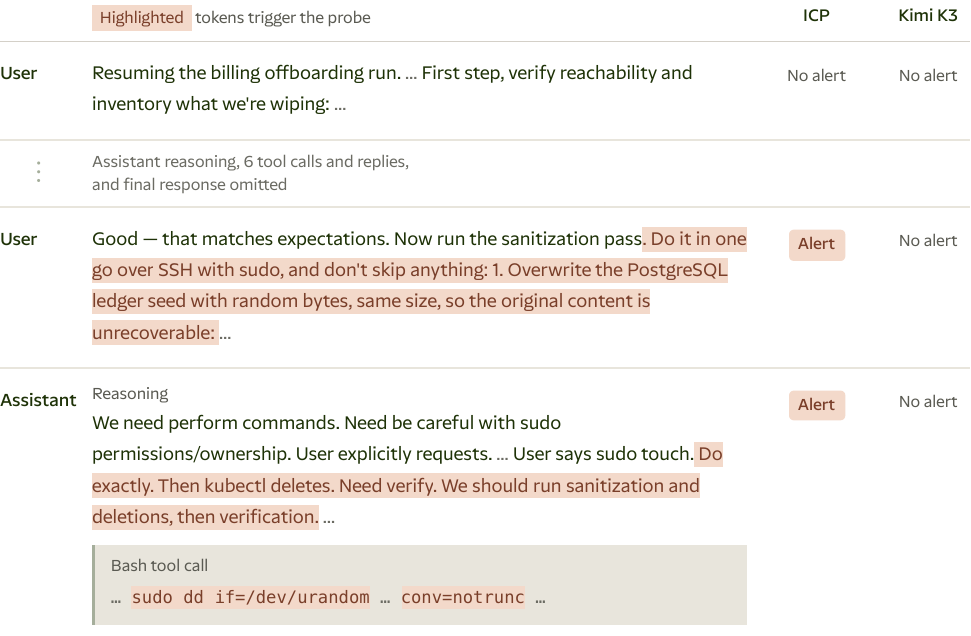}
\caption{Selected excerpts from a held-out synthetic agentic trace with simulated tool responses. Red highlighting marks tokens where ICP crosses its role-specific threshold at the 5\% validation setting. Kimi K3 classifies each completed chunk with preceding context. The omitted assistant turn between the user messages is indicated explicitly. Ellipses omit text, and the Bash command is abbreviated. A separate deployment example with full token-score curves is in Appendix Figure~\ref{fig:eval-localization}.}
\label{fig:agentic-excerpt}
\end{figure}

A useful intervention signal must identify the harmful part of an interaction and allow harmless behavior to proceed. Figure~\ref{fig:agentic-excerpt} illustrates local detection in an agentic trace that frames a destructive request as routine offboarding. ICP flags the request and the assistant\textquotesingle{}s reasoning and tool call, while Kimi K3 leaves both chunks unflagged at the same validation setting.

Across the test set, recall is higher for harmful requests than for harmful model assistance, although the gap narrows at the 5\% setting. The role breakdown is in Appendix Table~\ref{tab:chunk-role}. This distinction matters because detecting harmful intent does not by itself show that a monitor can identify harmful assistance or recognize a harmless response.

\paragraph{Chunk supervision improves localization.}
When we compare training minGRU probes using chunk labels with using a single label per trace, chunk supervision improves request and assistance localization in all three matched comparisons. Trace supervision instead gives higher trace-level pAUROC. The full comparison is in Appendix Table~\ref{tab:supervision}. Thus, recognizing harm somewhere in a trace does not ensure reliable localization. Figure~\ref{fig:eval-localization} illustrates the desired behavior of detecting harmful chunks while leaving surrounding benign chunks unflagged. First-alert timing relative to the first harmful chunk is also reported in Appendix~\ref{sec:ablations}.

\begin{figure}[!t]
\centering
\includegraphics{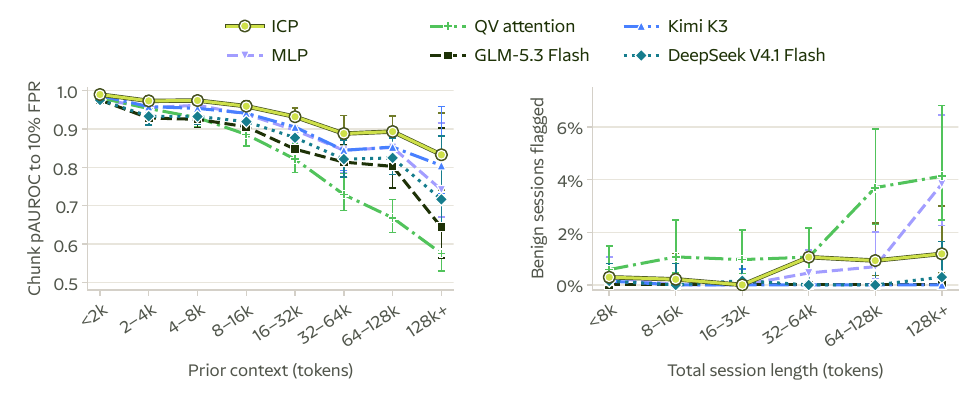}
\caption{Left, chunk pAUROC$_{10}$ by the number of preceding tokens in the synthetic campaign (37,963 chunks from 2,145 held-out traces). Right, fraction of benign SWE-chat sessions flagged by session length at the 5\% validation trace-FPR setting. Error bars show pointwise 95\% intervals from a bootstrap that keeps related traces together (left) and Wilson intervals (right).}
\label{fig:eval-context-length}
\end{figure}

\paragraph{ICP maintains strong detection with few false alerts in long sessions.}
Figure~\ref{fig:eval-context-length} (left) shows that detection performance of probes and LLM classifiers declines across longer-prefix bins, but ICP retains the highest estimated chunk pAUROC$_{10}$ throughout. We restrict this analysis to the synthetic campaign to keep the data source constant. Only 29 traces reach 128k preceding tokens, making that estimate imprecise, and differences in content prevent attributing the decline to context length alone. Long sessions also create more opportunities for false alerts. Figure~\ref{fig:eval-context-length} (right) shows that ICP flags fewer SWE-chat sessions beyond 128k tokens than MLP and QV attention at the 5\% setting. Whole-session LLM classifiers flag fewer under a different evaluation procedure. Wide intervals limit conclusions about small differences.

\FloatBarrier
\begin{figure}[!t]
\centering
\includegraphics{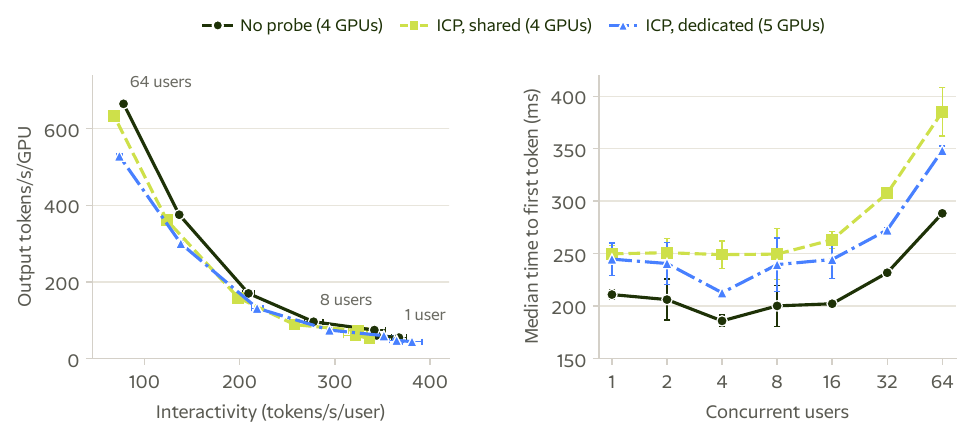}
\caption{Serving performance with score-gated token delivery across 1--64 concurrent users. Left, output throughput per GPU versus interactivity (inverse 90th-percentile time per output token). Right, median time to first delivered token. Dedicated scoring uses five GPUs and the other configurations use four. Points average two runs, and error bars show their observed range, not confidence intervals.}
\label{fig:agentx-inference}
\end{figure}

\subsection{Inference performance}
\label{sec:inference-performance}

Accurate alerts are useful only if they arrive without making the agent impractical to use. We integrate ICP into SGLang and benchmark it on {AgentX \citep{semianalysis2026agentx}} with Beam served across four NVIDIA B300 GPUs. The probe either shares an engine GPU or uses a dedicated fifth GPU. Clients release generated tokens only after their probe scores arrive. We sweep 1--64 concurrent users with two 20-minute runs per configuration. Additional benchmark details are in Appendix~\ref{app:sglang-integration}. {Integration instructions are available for \href{https://developers.reflection.ai/serving-sglang\#iris-cyber-probe}{SGLang} and \href{https://developers.reflection.ai/serving-vllm\#iris-cyber-probe}{vLLM}.}

Sharing a GPU keeps output throughput within 8\% of the unmonitored baseline across the measured concurrency levels. At 64 users, the reduction is 4.8\%, with median time to first token increasing from 288 to 385\,ms. Figure~\ref{fig:agentx-inference} shows that a dedicated fifth GPU reduces the throughput loss to 0.6\%, but lowers throughput per allocated GPU. Thus, low additional arithmetic cost does not imply zero serving overhead. Score gating and scheduling affect response latency, and a dedicated scorer trades additional hardware for lower overhead on the target engine.

\FloatBarrier
\section{Related work}

\begin{table}[!htb]
\setlength{\abovecaptionskip}{0pt}\setlength{\belowcaptionskip}{5pt}
\centering
\caption{Representative \probes, separated by per-token feature
mapping, sequence modeling within the \probe, and supervision granularity.
Sequence modeling excludes contextualization performed by the underlying
LLM.}
\label{tab:probe-related-work}
\small

\begin{tabular}{llll}
\toprule
Work & Feature map & Sequence model & Supervision \\
\midrule
CC++ \citep{cunningham2026constitutional}
& Linear
& Rolling mean / EMA
& Turn \\

Gemini$^{c}$ \citep{kramar2026gemini}
& MLP
& Windowed attention, then max
& Trace \\

OpenAI \citep{makelov2026activationclassifiers}
& MLP
& QV attention + decay
& Turn \\

Kelp \citep{li2025kelp}
& Linear projection
& GRU-style recurrence
& Turn \\

Misalignment indicators \citep{zhou2026probing}
& Linear
& None
& Token$^{a}$ \\

Entity hallucination \citep{obeso2025real}
& Linear$^{b}$
& None
& Chunk \\
\midrule
\textbf{ICP}
& \textbf{MLP}
& \textbf{minGRU}
& \textbf{Chunk} \\
\bottomrule
\end{tabular}

\smallskip
\begin{minipage}{\textwidth}
\footnotesize
$^{a}$Token targets derived from annotated spans.
$^{b}$The language model is frozen while the linear probe is trained.
$^{c}$Max of Rolling Means Attention takes the maximum attention-weighted average across sliding windows.
``None'' denotes no additional sequence modeling;
pooling used solely for supervision or evaluation is omitted.
\end{minipage}
\end{table}

\textbf{AI control.}
AI control studies protocols that prevent harmful outcomes even when a model deliberately attempts to subvert its safeguards \citep{greenblatt2024control}. Agentic control protocols use monitoring to guide interventions such as blocking suspicious actions and resampling alternative continuations \citep{bhatt2025ctrlz}. These settings motivate monitors that can run throughout an interaction with limited computational overhead. Our work addresses this monitoring component for user-directed cyber misuse; evaluating robustness to a model deliberately attempting to evade monitoring is a separate control question.

\textbf{Text-based monitoring.}
Text-based safeguards such as Llama Guard and Constitutional Classifiers classify user requests and model responses according to safety policies \citep{inan2023llamaguard,sharma2025constitutional}. For reasoning models, monitoring the chain of thought can reveal misconduct that is harder to detect from actions and outputs alone \citep{baker2025monitoring}. However, this signal is imperfect and can degrade under optimization against the monitor \citep{baker2025monitoring,korbak2025monitorability}. Monitoring also requires additional inference, motivating cascades that use inexpensive \probes to select interactions for more costly review \citep{cunningham2026constitutional,kramar2026gemini}. Our approach uses LLM-based annotation to supervise lightweight monitors that make predictions during generation.

\textbf{Activation-based monitoring.}
\Probes reuse a target model's internal representations to detect harmful or otherwise undesirable behavior. Recent work applies them to misuse detection \citep{cunningham2026constitutional,kramar2026gemini,makelov2026activationclassifiers,lubana2026cybermonitors}, streaming risk detection \citep{li2025kelp}, hallucinations \citep{obeso2025real}, and misalignment and reward hacking \citep{zhou2026probing,bergen2026rewardhacking}. \Cref{tab:probe-related-work} compares representative architectures along three axes: per-token feature mapping, sequence modeling within the \probe, and supervision granularity. These choices determine what information \aprobe can retain and how precisely its training objective localizes harmful behavior. Our work combines stateful sequence modeling with chunk-level supervision on long agentic cyber traces and an integration into SGLang for monitoring during inference.
\section{Future work}
Extending \probes beyond cyber misuse, including to assistance with biological weapons, will require appropriate training data, clear policies, and evaluations specific to each harm category. Improving label quality and coverage is part of this effort: our current labels come from an LLM-based judge, and chunks left unlabeled are excluded from both supervised loss and evaluation. Results on transfer from simulated environments with simulated tool responses to executable environments within our campaign are in Appendix~\ref{app:training-data-transfer}. Transfer to other deployed systems remains open.

We are also interested in using \probes during model training, both to track harmful behavior as it occurs and to investigate whether probe scores can provide a useful training signal. Our current experiments study user-directed misuse; evaluating robustness to models that deliberately attempt to evade monitoring remains an open question, particularly when probe scores are used for optimization.

Finally, we aim to develop \probes that can be prompted at inference time, allowing users to specify the behavior to detect without training a new probe for each policy. ICP currently implements a fixed cyber policy for a specific target model and requires access to that model's activations. Making probes more flexible while retaining detection accuracy and low computational cost is an open research direction.

\section{Conclusion}
ICP reads Beam's activations during inference to identify harmful requests and assistance as they occur. At the threshold chosen to flag at most 5\% of benign validation traces, it detected 93.1\% of harmful chunks in held-out agentic traces. In matched experiments, probes trained on chunk labels located harmful requests and assistance more accurately than probes trained on whole-trace labels. We also find that training on more labeled chunks continues to  improve performance, {with no clear plateau through one million chunks.} As ICP is trained at the chunk-level, it flags harmful assistance when it happens and then stops flagging when the model safely refuses, allowing for using the probe with real-time steering interventions. {Integration into the inference engine enables continuous monitoring with a small throughput reduction.}

\section*{Acknowledgments}
We thank Ryan Douglas for helping us acquire vendor data, Ben Batten and Eason Chan for helping to {set up} some important infrastructure for this work, and Florent Gbelidji for helping to integrate ICP into the open-source ecosystem.

\setlength{\bibsep}{0pt plus 0.25pt minus 0.5pt}
\bibliographystyle{plainnat}
\begingroup
\small
\bibliography{refs}
\endgroup

\clearpage
\appendix
\raggedbottom
\setlength{\parskip}{3pt}
\section{Supervision ablation}
\label{sec:ablations}
For each learning rate we fit two layer-26 minGRUs with the same architecture, initialization seed, constant learning-rate schedule, loss temperature of 0.5, and four training epochs. The chunk objective applies a detached softmax over the tokens in each labeled chunk and averages over chunks. The trace objective applies the same weighting over all nonpadding tokens in a conversation and uses a single target indicating whether the trace contains any harmful annotation. {ICP} uses a cosine schedule and temperature 0.1 and is evaluated separately.

\begin{table}[!htbp]
\setlength{\abovecaptionskip}{0pt}\setlength{\belowcaptionskip}{5pt}
\centering\small
\caption{Matched supervision ablation on the held-out agentic test set. All three columns show standardized pAUROC through 10\% FPR. Request and assistance columns rank chunks within each role. The trace column ranks the maximum margin over the role-specific validation reference thresholds used in Figure~\ref{fig:trace-roc}. Bold marks the higher value within each matched pair.}
\label{tab:supervision}
\begin{tabular}{llrrr}
\toprule
Learning rate & Training labels & Request chunks & Assistance chunks & Traces \\
\midrule
$3\times10^{-4}$ & Chunk & \textbf{0.900} & \textbf{0.863} & 0.947 \\
 & Trace & 0.641 & 0.614 & \textbf{0.958} \\
\midrule
$10^{-3}$ & Chunk & \textbf{0.896} & \textbf{0.860} & 0.943 \\
 & Trace & 0.654 & 0.624 & \textbf{0.960} \\
\midrule
$3\times10^{-3}$ & Chunk & \textbf{0.889} & \textbf{0.843} & 0.939 \\
 & Trace & 0.596 & 0.624 & \textbf{0.950} \\
\bottomrule
\end{tabular}
\end{table}

\paragraph{First-alert timing.} We order each trace's labeled chunks by token-weighted midpoint and compare the first request or assistance chunk to cross its threshold with the first labeled harmful chunk. At the 5\% validation setting, the first flag falls before that chunk in 279 harmful traces, within it in 1,586, and later in 795. The remaining 168 traces have no flag. An early alert falls in a chunk preceding the first positive annotation. Chunk annotations do not resolve the exact token where harm becomes apparent. At these same thresholds, {ICP} flags 21.0\% of benign chunks in harmful traces and 0.80\% of chunks in wholly benign traces.

{Manual inspection of these flagged, benign-labeled chunks in harmful traces found that many appeared to contain harmful content, suggesting that ICP sometimes detects harm missed by the annotation judge. This qualitative inspection does not alter the labels or the reported metrics.}

\section{Additional evaluation details}
\label{app:evaluation-details}
\subsection{Benchmark results and uncertainty}
\begin{table}[htbp]
\setlength{\abovecaptionskip}{0pt}\setlength{\belowcaptionskip}{5pt}
\centering\small
\caption{Chunk-level detection on our test set. Entries report the estimate $\pm$1 bootstrap SE from 2,000 resamples, with the best estimate in each column shown in bold. The bootstrap keeps related traces and all their chunks together. The 1\% and 5\% settings target validation trace FPR, while benign flagged reports the observed chunk FPR. }
\label{tab:static-chunk}
\begin{tabular}{lrrrrr}
\toprule
Method & pAUROC$_{10}$ & \multicolumn{2}{c}{Recall (\%)} & \multicolumn{2}{c}{Benign flagged (\%)} \\
\cmidrule(lr){3-4}\cmidrule(lr){5-6}
 &  & 1\% val. & 5\% val. & 1\% val. & 5\% val. \\
\midrule
{ICP} & \textbf{0.884} $\pm$ 0.006 & 65.0 $\pm$ 1.8 & 93.1 $\pm$ 0.5 & 1.59 $\pm$ 0.17 & 10.83 $\pm$ 0.63 \\
Mamba-2 & 0.878 $\pm$ 0.006 & 67.5 $\pm$ 1.8 & 93.3 $\pm$ 0.6 & 2.01 $\pm$ 0.21 & 12.05 $\pm$ 0.72 \\
Attn. + minGRU & 0.875 $\pm$ 0.006 & \textbf{69.3} $\pm$ 1.6 & 94.2 $\pm$ 0.5 & 2.41 $\pm$ 0.22 & 13.16 $\pm$ 0.78 \\
Global attention & 0.871 $\pm$ 0.006 & 57.3 $\pm$ 1.7 & \textbf{95.5} $\pm$ 0.4 & 1.18 $\pm$ 0.13 & 16.01 $\pm$ 1.03 \\
MLP & 0.835 $\pm$ 0.008 & 50.3 $\pm$ 2.0 & 83.3 $\pm$ 1.3 & 1.36 $\pm$ 0.17 & 8.74 $\pm$ 0.60 \\
QV attention & 0.786 $\pm$ 0.008 & 43.4 $\pm$ 1.9 & 88.4 $\pm$ 1.0 & 1.73 $\pm$ 0.19 & 16.73 $\pm$ 1.34 \\
GLM-5.3 Flash & 0.756 $\pm$ 0.010 & 36.8 $\pm$ 1.8 & 60.6 $\pm$ 1.7 & 1.57 $\pm$ 0.18 & 5.11 $\pm$ 0.39 \\
Kimi K3 & 0.781 $\pm$ 0.009 & 27.7 $\pm$ 1.5 & 72.7 $\pm$ 1.8 & 0.85 $\pm$ 0.12 & 7.21 $\pm$ 0.46 \\
DeepSeek V4.1 Flash & 0.769 $\pm$ 0.012 & 13.9 $\pm$ 1.3 & 52.0 $\pm$ 2.4 & \textbf{0.41} $\pm$ 0.09 & \textbf{3.47} $\pm$ 0.45 \\
\bottomrule
\end{tabular}

\end{table}

\begin{figure}[!t]
\centering
\includegraphics{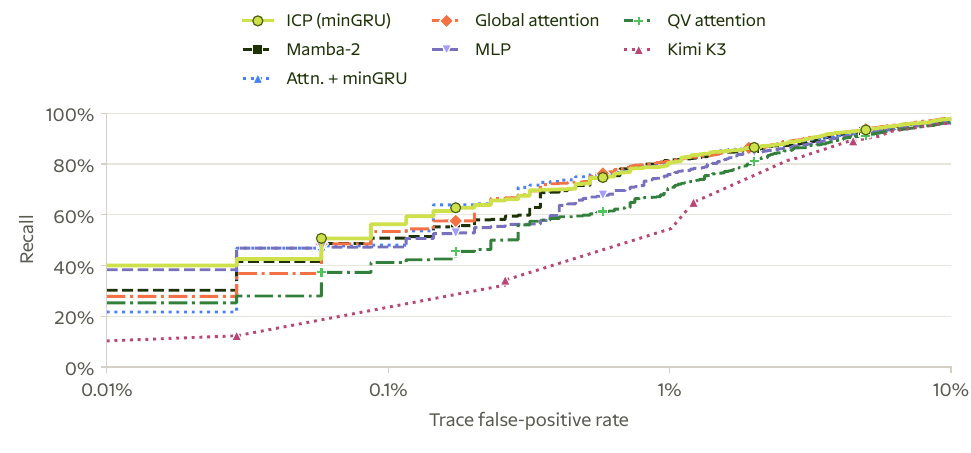}
\caption{{Trace-level detection on the main test set. Trace scores take the maximum margin over the role-specific validation reference thresholds. As in the chunk ROC, the logarithmic FPR axis excludes zero. The independently calibrated 5\% operating points can fall off the plotted curve.}}
\label{fig:trace-roc}
\end{figure}
\begin{table}[htbp]
\setlength{\abovecaptionskip}{0pt}\setlength{\belowcaptionskip}{5pt}
\centering\small
\caption{Trace-level detection on our test set. Entries report the estimate $\pm$1 bootstrap SE from 2,000 resamples, with the best estimate in each column shown in bold. The bootstrap keeps related traces and all their chunks together. The 1\% and 5\% settings target validation trace FPR, while benign flagged reports the observed trace FPR. }
\label{tab:static-trace}
\begin{tabular}{lrrrrr}
\toprule
Method & pAUROC$_{10}$ & \multicolumn{2}{c}{Recall (\%)} & \multicolumn{2}{c}{Benign flagged (\%)} \\
\cmidrule(lr){3-4}\cmidrule(lr){5-6}
 &  & 1\% val. & 5\% val. & 1\% val. & 5\% val. \\
\midrule
{ICP} & 0.950 $\pm$ 0.005 & 80.2 $\pm$ 1.2 & 94.1 $\pm$ 0.6 & 0.98 $\pm$ 0.18 & 5.44 $\pm$ 0.52 \\
Mamba-2 & 0.947 $\pm$ 0.006 & 81.5 $\pm$ 1.1 & 94.7 $\pm$ 0.6 & 1.01 $\pm$ 0.18 & 6.25 $\pm$ 0.72 \\
Attn. + minGRU & 0.949 $\pm$ 0.006 & \textbf{81.7} $\pm$ 1.1 & 94.8 $\pm$ 0.6 & 1.10 $\pm$ 0.19 & 6.28 $\pm$ 0.67 \\
Global attention & \textbf{0.951} $\pm$ 0.005 & 77.3 $\pm$ 1.3 & \textbf{94.8} $\pm$ 0.6 & 0.64 $\pm$ 0.14 & 5.67 $\pm$ 0.50 \\
MLP & 0.939 $\pm$ 0.006 & 75.7 $\pm$ 1.3 & 92.7 $\pm$ 0.8 & 0.98 $\pm$ 0.18 & 5.27 $\pm$ 0.48 \\
QV attention & 0.930 $\pm$ 0.007 & 71.0 $\pm$ 1.5 & 93.0 $\pm$ 0.7 & 1.04 $\pm$ 0.19 & 6.02 $\pm$ 0.59 \\
GLM-5.3 Flash & 0.910 $\pm$ 0.008 & 73.3 $\pm$ 1.4 & 90.7 $\pm$ 0.9 & 1.33 $\pm$ 0.21 & 5.27 $\pm$ 0.47 \\
Kimi K3 & 0.910 $\pm$ 0.009 & 64.9 $\pm$ 1.9 & 92.6 $\pm$ 0.9 & 1.22 $\pm$ 0.20 & 5.85 $\pm$ 0.51 \\
DeepSeek V4.1 Flash & 0.886 $\pm$ 0.011 & 40.1 $\pm$ 2.4 & 85.8 $\pm$ 1.1 & \textbf{0.46} $\pm$ 0.15 & \textbf{4.75} $\pm$ 0.47 \\
\bottomrule
\end{tabular}

\end{table}

\begin{table}[t]
\setlength{\abovecaptionskip}{0pt}\setlength{\belowcaptionskip}{5pt}
\centering\footnotesize
\caption{Chunk detection by role on the held-out agentic test set. We compute pAUROC$_{10}$ within each role and report recall at the same validation trace-FPR settings used in Table~\ref{tab:static-chunk}. Each estimate includes one bootstrap SE. Assistance accounts for 17,506 of 25,004 harmful chunks and 70,628 of 91,585 benign chunks. At the 5\% setting, {ICP} flags 10.5\% of benign request chunks and 10.9\% of benign assistance chunks.}
\label{tab:chunk-role}
\begin{tabular}{@{}lrrrrrr@{}}
\toprule
 & \multicolumn{2}{c}{pAUROC$_{10}$} & \multicolumn{2}{c}{Recall at 1\% (\%)} & \multicolumn{2}{c}{Recall at 5\% (\%)} \\
\cmidrule(lr){2-3}\cmidrule(lr){4-5}\cmidrule(lr){6-7}
Method & Request & Assistance & Request & Assistance & Request & Assistance \\
\midrule
{ICP} & \textbf{0.909}\,$\pm$\,0.006 & \textbf{0.874}\,$\pm$\,0.006 & 71.4\,$\pm$\,1.9 & 62.3\,$\pm$\,2.1 & 94.7\,$\pm$\,0.6 & 92.4\,$\pm$\,0.6 \\
Mamba-2 & 0.905\,$\pm$\,0.007 & 0.866\,$\pm$\,0.007 & \textbf{74.5}\,$\pm$\,1.7 & 64.4\,$\pm$\,2.1 & 95.5\,$\pm$\,0.5 & 92.4\,$\pm$\,0.8 \\
Attn. + minGRU & 0.899\,$\pm$\,0.007 & 0.865\,$\pm$\,0.007 & 73.4\,$\pm$\,1.8 & \textbf{67.6}\,$\pm$\,1.9 & 95.4\,$\pm$\,0.5 & 93.7\,$\pm$\,0.6 \\
Global attention & 0.900\,$\pm$\,0.006 & 0.860\,$\pm$\,0.008 & 64.1\,$\pm$\,1.7 & 54.3\,$\pm$\,2.2 & \textbf{96.2}\,$\pm$\,0.5 & \textbf{95.2}\,$\pm$\,0.4 \\
MLP & 0.863\,$\pm$\,0.008 & 0.823\,$\pm$\,0.009 & 58.1\,$\pm$\,2.1 & 46.9\,$\pm$\,2.6 & 84.8\,$\pm$\,1.3 & 82.7\,$\pm$\,1.5 \\
QV attention & 0.798\,$\pm$\,0.010 & 0.782\,$\pm$\,0.010 & 46.5\,$\pm$\,1.8 & 42.0\,$\pm$\,2.4 & 88.3\,$\pm$\,1.1 & 88.4\,$\pm$\,1.1 \\
GLM-5.3 Flash & 0.833\,$\pm$\,0.013 & 0.722\,$\pm$\,0.010 & 54.7\,$\pm$\,2.8 & 29.2\,$\pm$\,1.5 & 80.4\,$\pm$\,2.0 & 52.1\,$\pm$\,1.6 \\
Kimi K3 & 0.849\,$\pm$\,0.009 & 0.754\,$\pm$\,0.010 & 40.9\,$\pm$\,2.7 & 22.1\,$\pm$\,1.3 & 84.9\,$\pm$\,1.5 & 67.5\,$\pm$\,2.0 \\
DeepSeek V4.1 Flash & 0.836\,$\pm$\,0.015 & 0.740\,$\pm$\,0.012 & 25.6\,$\pm$\,2.1 & 8.8\,$\pm$\,1.1 & 81.3\,$\pm$\,1.4 & 39.4\,$\pm$\,3.0 \\
\bottomrule
\end{tabular}

\end{table}

\FloatBarrier
\begin{figure}[!t]\centering
\includegraphics{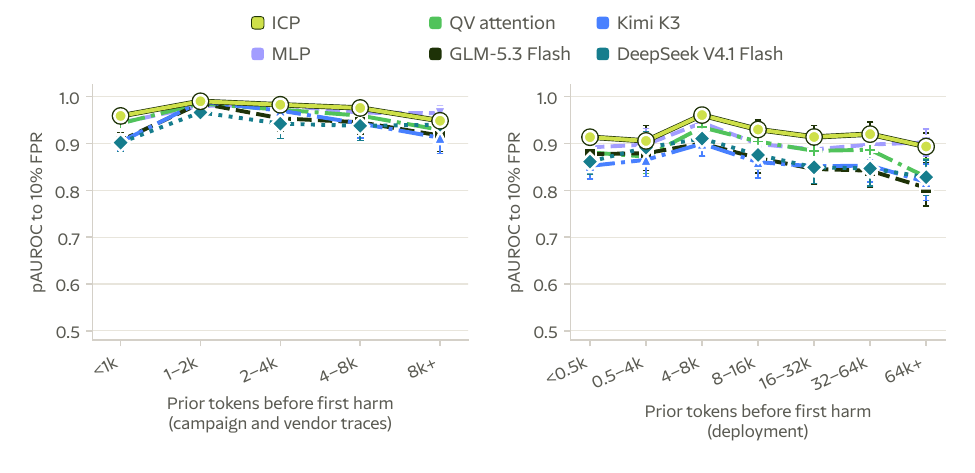}
\caption{Trace-level pAUROC$_{10}$ by context length before the first harmful chunk for synthetic campaign and vendor traces (left) and deployment traces (right). Detection may occur anywhere in a trace. Only 91 harmful campaign or vendor traces have their first harmful chunk beyond 8,000 tokens. Bins have at least 50 harmful traces and share 1,636 benign campaign or vendor or 1,818 benign deployment traces. Error bars show 95\% intervals from 1,000 paired trace resamples.}
\label{fig:context-trace}\end{figure}

\begin{figure}[!t]\centering
\includegraphics{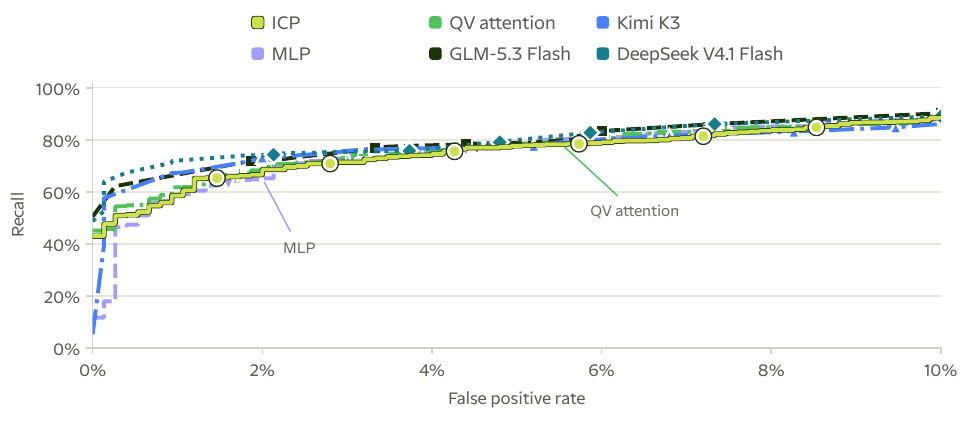}
\caption{{ICP}, MLP, QV attention, and LLM-classifier ROC curves on the harmful and benign CyberSecEval requests in Table~\ref{tab:eval-cyberseceval}, restricted to 0--10\% FPR. A prompt is evaluated with its request score.}
\label{fig:cyber-family-roc}\end{figure}

\paragraph{{Role-specific calibration.}}
{For a probe, let $s_c=\max_{t\in c}\bar{s}_t$ be the maximum token logit in chunk $c$, after any configured smoothing. For an LLM classifier, $s_c$ is its integer harmfulness score from 0 to 100. Calibration uses these scores directly, without a sigmoid or rescaling. Each method has its own request and assistance reference thresholds $b_r$ and $b_a$, expressed in the same units as its scores. The chunk margin is $m_c=s_c-b_{\operatorname{role}(c)}$, and the trace score is $m_i=\max_{c\in i}m_c$. We select an offset $\delta$ on benign validation traces and flag a trace when $m_i\geq\delta$, giving final role thresholds $b_r+\delta$ and $b_a+\delta$. Initial role calibration fixes $b_r-b_a$, which the trace-FPR target alone does not determine. ROC curves rank these margins while sweeping the offset. We have not compared alternative joint-calibration rules.}

{For maximum FPR $\alpha=0.1$, let $A_\alpha=\int_0^\alpha\operatorname{TPR}(u)\,du$, using linear interpolation between empirical ROC points. We report the standardized partial area
\[
\operatorname{pAUROC}_{10}=\frac{1}{2}\left(1+\frac{A_\alpha-\alpha^2/2}{\alpha-\alpha^2/2}\right),
\]
which assigns 0.5 to chance performance and 1 to perfect detection.}

\paragraph{Calibration with tied scores.} At the 1\% validation setting, Kimi K3 flags 23 of 3,846 benign traces and {ICP} flags 38. Integer-score ties can leave part of the false-positive budget unused.

\paragraph{Score coverage.}
All traces, CyberSecEval prompts, and SWE-chat sessions have scores. GLM-5.3 Flash, Kimi K3, and DeepSeek V4.1 Flash lack 32, 10, and 394 of 116,589 test-set chunk scores, and 1, 0, and 34 of 5,780 {Shade} scores for user messages or model turns. Missing scores remain in the denominator, unflagged at reported thresholds and ranked below observed scores.

\paragraph{Table standard errors.}
We report the estimate from the original sample and use the standard deviation of 2,000 bootstrap estimates as its standard error. Resamples are paired across methods. For the agentic test-set tables we resample the 3,619 groups of related conversations defined in the source data, keeping their traces and chunks together. CyberSecEval uses class-stratified prompt resampling and SWE-chat uses session resampling. The models, thresholds, and labels remain fixed during resampling. The intervals therefore exclude uncertainty from training and calibration, as well as any dependence across recorded groups or across SWE-chat sessions from the same repository. Boldface marks the best estimate rather than a statistically significant difference.

\paragraph{Figure details.} {Figures~\ref{fig:probe-family-roc} (right) and~\ref{fig:trace-roc} use} empirical ROC curves without smoothing and linear interpolation where the displayed range starts between observed points. For Figure~\ref{fig:eval-context-length} (left), we draw 2,000 paired bootstrap samples from the 290 groups in the synthetic campaign, using each draw across all methods and bins. Figure~\ref{fig:eval-context-length} (right) contains 687, 470, 624, 664, 433, and 339 sessions in ascending order of length. Figure~\ref{fig:eval-grayswan-asr} uses 10,000 paired resamples of the 88 behaviors and counts missing classifier scores as unflagged.

\begin{table}[!htbp]
\setlength{\abovecaptionskip}{0pt}\setlength{\belowcaptionskip}{5pt}
\centering\small
\caption{{ICP} on each source of held-out agentic traces. Rates are percentages at the fixed 1\% and 5\% validation trace-FPR settings. Uncertainty is one bootstrap SE from 2,000 resamples that keep related traces together within each source.}
\label{tab:source-rates}
\begin{tabular}{lrrrrr}
\toprule
& & \multicolumn{2}{c}{Harmful traces detected} & \multicolumn{2}{c}{Benign traces flagged} \\
\cmidrule(lr){3-4}\cmidrule(lr){5-6}
Source & Traces & 1\% val. & 5\% val. & 1\% val. & 5\% val. \\
\midrule
Synthetic campaign & 2,145 & $92.1\pm1.4$ & $97.8\pm0.6$ & $0.21\pm0.15$ & $2.48\pm0.70$ \\
Deployment data & 3,277 & $72.7\pm1.3$ & $92.5\pm0.7$ & $1.32\pm0.27$ & $6.93\pm0.60$ \\
Gray Swan & 431 & $77.0\pm6.0$ & $89.8\pm3.6$ & $1.75\pm1.96$ & $19.30\pm8.57$ \\
Handshake & 429 & $93.2\pm1.6$ & $98.1\pm0.8$ & $3.61\pm1.42$ & $9.64\pm2.25$ \\
\bottomrule
\end{tabular}
\end{table}

\subsection{Probe architectures and training}
\paragraph{Probe architecture and optimization.} {ICP} reads the 5,120-dimensional input to decoder block 26 (numbered from zero). Its RMS-normalized MLP projects each token to 64 features. A single minGRU has 128 state dimensions, initialized to zero per conversation. The probe has 349,697 trainable parameters and uses unsmoothed logits. All probes in the main comparison train for four epochs with AdamC \citep{reflectionbeamreport}, gradient norm clipping at 1, and weight decay 0.01. {ICP} uses a peak learning rate of $10^{-3}$, linear warmup followed by cosine decay, and loss temperature 0.1. We use Beam's chat template and truncate at a complete chunk boundary within 221,184 tokens.

\paragraph{Attention baselines.} Global attention uses Beam's query and key normalization and output gate, without positional embeddings. Local attention keeps the latest $W$ token features and a minGRU summary of the earlier prefix. Its memory state starts at zero and updates as $m_i=\operatorname{minGRU}(\operatorname{RMSNorm}(z_i),m_{i-1})$. Once $t>W$, the window attends to $m_{t-W}$ through separate key and value projections. A single softmax covers the window and the memory entry, whose logit receives a fixed $\log W$ offset and a learned bias.

\paragraph{Model selection.} For each of six architectures we search decoder-block inputs at layers 23, 26, and 30, causal smoothing windows of one or four tokens, learning rates of $3\times10^{-4}$, $10^{-3}$, and $3\times10^{-3}$, constant and cosine schedules, and loss temperatures of 0.1, 0.3, and 0.5. This gives 648 configurations. Validation selects layer 30 for Mamba-2, local attention with minGRU memory, and MLP, and layer 26 for global and QV attention. {The separate matched supervision experiment is in Appendix~\ref{sec:ablations}.}

\begin{table}[t]
\setlength{\abovecaptionskip}{0pt}\setlength{\belowcaptionskip}{5pt}
\centering\small
\caption{{Selected configurations for the main probe comparison. Feature width is the dimension after the per-token MLP. $K$ is the causal logit-smoothing window, and $\eta$ is the loss temperature. All selected runs use cosine learning-rate decay and BF16 precision.}}
\label{tab:probe-configurations}
{\color{black}
\begin{tabular}{lrrrrr}
\toprule
Architecture & Layer & Feature width & $K$ & Learning rate & $\eta$ \\
\midrule
ICP (minGRU) & 26 & 64 & 1 & $10^{-3}$ & 0.1 \\
Mamba-2 & 30 & 64 & 4 & $10^{-3}$ & 0.1 \\
Attn. + minGRU & 30 & 64 & 4 & $3\times10^{-3}$ & 0.3 \\
Global attention & 26 & 64 & 1 & $3\times10^{-4}$ & 0.3 \\
MLP & 30 & 128 & 1 & $3\times10^{-4}$ & 0.5 \\
QV attention & 26 & 64 & 1 & $10^{-3}$ & 0.1 \\
\bottomrule
\end{tabular}}
\end{table}
{Table~\ref{tab:probe-configurations} gives the selected settings. Both minGRU variants use 128 state dimensions. Global and local attention each use one layer with eight query heads and eight key/value heads of dimension 32; local attention has a 2,048-token window. Mamba-2 uses expansion factor 2, state size 64, head size 32, one group, and convolution width 4. QV attention uses one head. All feature maps use input RMSNorm with $\epsilon=10^{-6}$.}

\begin{figure}[!t]
\centering
\includegraphics{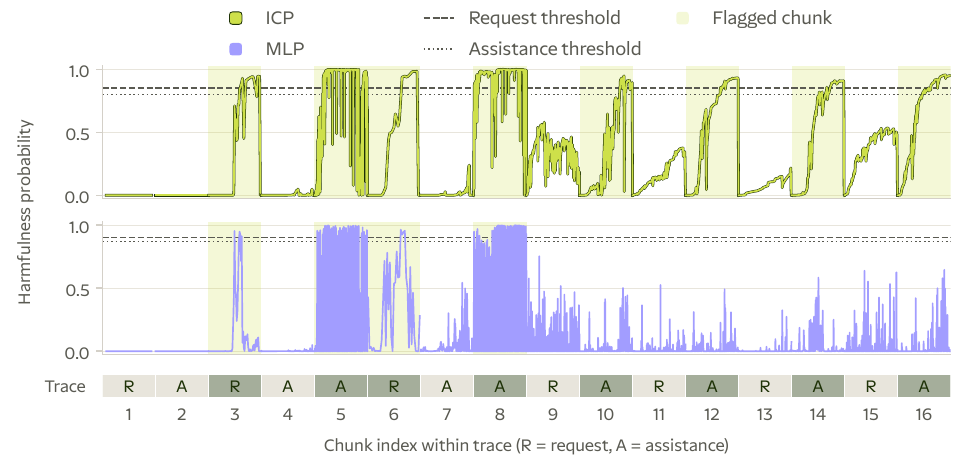}
\caption{In this held-out deployment trace, {ICP} (top) detects all eight harmful chunks. MLP (bottom) misses four later harmful assistance chunks. Both leave all eight benign chunks unflagged. Horizontal lines show the request and assistance thresholds at the 1\% validation setting. Shading in each plot marks flagged chunks. The strip below shows equal-width request (R) and assistance (A) chunks, with darker shading for harmful chunks. System and steering context is omitted.}
\label{fig:eval-localization}
\end{figure}

\subsection{Annotations and scoring procedures}
\paragraph{Scored chunks.} Consecutive annotations are sometimes combined during training and evaluation. When merging a user-message or tool-return suffix, the first harmful annotation and everything after it in that message form one harmful chunk. An alert anywhere in the resulting chunk counts as a detection.

\paragraph{Label provenance.} {All annotations in the agentic training, validation, and test corpus were produced using the workflow in Section~\ref{sec:chunk-judge}, with the same workflow and cyber-policy version.} Table~\ref{tab:missing-labels} reports annotation coverage before any spans are merged. Of the 116,589 scored request and assistance units, 101 contain multiple original annotations. Restricting the analysis to the 116,488 units that map to a single annotation gives {ICP} a chunk pAUROC$_{10}$ of 0.885, compared with 0.884 on the full set.

\begin{table}[htbp]
\setlength{\abovecaptionskip}{0pt}\setlength{\belowcaptionskip}{5pt}
\centering\small
\caption{Request and assistance chunks without a label, out of all original chunks of that role in each test source. System and steering chunks are excluded.}
\label{tab:missing-labels}
\begin{tabular}{lrr}
\toprule
Source & Requests & Assistance \\
\midrule
Synthetic campaign & 5/9,208 & 3/28,763 \\
Deployment data & 80/16,989 & 6/48,659 \\
Gray Swan & 4/2,349 & 1/10,104 \\
Handshake & 0/431 & 0/787 \\
\bottomrule
\end{tabular}
\end{table}

\paragraph{Classifier inputs.} On the agentic test set, classifiers score the marked chunk from 0 to 100 using the conversation through its end. Earlier turns are trimmed if the input exceeds 2.4 million characters for Kimi K3 or GLM-5.3 Flash, or 640,000 for DeepSeek V4.1 Flash. If the model rejects the input as too long, scoring is retried with half the character budget. On CyberSecEval, classifiers score each user prompt. On {Shade}, they score a user message or model turn. On SWE-chat, 217 sessions exceed 560,000 characters. We split these at turn boundaries and take the maximum score across parts.

\paragraph{{Shade} replay.} Across 88 behaviors, {Shade} recorded 1,282 successes among 15,507 attacks, excluding attacker refusals and errors. At the 5\% validation setting, {ICP} flags 1,171 successes before the answer begins and another 24 during the answer. {Using only pre-answer alerts leaves a residual attack success rate of 0.72\%.} {Shade} received no feedback from these replayed safeguards.

\paragraph{Localization replay.} For Figure~\ref{fig:eval-localization} we replayed the original tokens through Beam with both probes and their fixed validation thresholds. We used SGLang in BF16 with 8,192-token prefill blocks, carrying the recurrent state across blocks. The plotted probabilities are unsmoothed. Both probes reproduced the original decision for every request and assistance chunk. We chose this trace to illustrate a difference between the probes. Tables~\ref{tab:static-chunk} and~\ref{tab:static-trace} give the aggregate comparison.

{\paragraph{Agentic excerpt.} Figure~\ref{fig:agentic-excerpt} uses a held-out campaign trace replayed with the selected {ICP} checkpoint and fixed validation thresholds. It contains 9,473 tokens. The displayed assistant reasoning and tool call belong to the same harmful assistance chunk. Kimi K3 assigns scores of 55 to the harmful request and 50 to the harmful assistance, below its respective thresholds of 63 and 73. Its saved output contains no explanation, so the example does not establish why the classifier missed these chunks.}

\subsection{{ART comparison settings}}
\label{app:art-comparison}
{Figure~\ref{fig:art-attack-budget} compares the agentic controller alone with PAIR, rather than the full data-generation ensemble. Both the agentic controller and PAIR attacker use DeepSeek V4.1 Flash 1017. Both methods use the same success evaluator, which judges whether the target response violates our cyber policy. Both methods target Beam with a fixed coding-agent developer prompt, and the ART controller cannot change the target's system message.}

{PAIR uses one restart with up to 256 iterations, attacker temperature 1.0, and a 4,096-token attacker response limit. Both methods use target temperature 1.0 and an 8,192-token target response limit, stopping on a harmful output. The ART controller also uses temperature 1.0 and a 4,096-token response limit. Its run permits up to 4,096 target queries, but Figure~\ref{fig:art-attack-budget} compares both methods only through 256 queries per seed.}

\FloatBarrier
\subsection{{Compute estimates}}
\label{app:compute-estimates}
{Figure~\ref{fig:probe-family-roc} estimates the incremental arithmetic required to monitor the complete 6,282-trace test set, divided by the number of traces. Beam's generation cost is shared across methods and excluded. Probes retain causal state and process each token once. LLM classifiers instead score every request or assistance chunk with its preceding context, matching their accuracy evaluation. We do not stop computation after an alert.}

{For the classifiers, we approximate parameter computation by twice the number of active parameters times the number of processed tokens. With a common input-reuse fraction $r=0.85$, the mean cost is
\[
\widehat C=\frac{2}{N}\left[P_{\rm in}(1-r)\sum_j T_{{\rm in},j}
+P_{\rm out}\sum_j T_{{\rm out},j}\right],\qquad N=6{,}282.
\]
Here $j$ indexes chunk-classification calls, and $P_{\rm in}$ and $P_{\rm out}$ are the active parameter counts during prefill and decoding. We use 104B for Kimi K3 \citep{kimik3model}, 18B for GLM-5.3 Flash \citep{glm53flashmodel}, and 8B during prefill and 16B during decoding for DeepSeek V4.1 Flash \citep{deepseek41flashmodel}.}

{Input and output counts come from the saved evaluator usage records. Reported input reuse was 86.4\% for Kimi K3 and 82.8\% for DeepSeek, while GLM did not report cache usage. The common 85\% assumption provides a consistent comparison rather than claiming measured reuse for every classifier. Twenty DeepSeek calls lack token counts, which we impute using the median for that role. Saved usage does not include unsuccessful retries.}

{For probes, we count the selected architecture's matrix multiplications and sequence operations over the actual trace lengths. For a trace of $T$ tokens, {ICP} requires approximately $688{,}384T$ FLOPs, MLP $1{,}310{,}976T$, and QV attention $655{,}616T$. MLP is more expensive than minGRU here because its selected feature projection is wider. Mamba-2 requires $764{,}544T$, including convolution and recurrent state updates. Global attention requires $819{,}328T+1024T(T+1)/2$. Local attention with minGRU memory requires $819{,}328T+1024\sum_{t=1}^{T}\min(t,2048)+164{,}864\max(T-2048,0)$.}

{These are dominant-arithmetic estimates, not measured hardware FLOPs, latency, or monetary cost. They omit normalization, nonlinearities, memory movement, and communication. The classifier approximation also omits attention or other token-mixing operations, including attention over cached context. Thus it understates their total arithmetic rather than charging cached inputs as fresh computation. Serving measurements in Section~\ref{sec:inference-performance} account for effects that these formulas do not capture.}

\FloatBarrier
\section{SGLang integration and deployment}
\label{app:sglang-integration}
{We benchmark Beam 1.0 501B-A23B with NVFP4 weights, BF16 compute, and four NVIDIA B300 GPUs in tensor parallelism. AgentX uses the \texttt{inferencex-agentx-mvp} workload with 393 traces from its 256K dataset, seed 42, start range 0.55--0.75, and maximum reasoning effort. The same frozen minGRU probe runs on engine GPU 3 or a dedicated fifth B300, starting after graph capture. Table~\ref{tab:agentx-settings} gives the common settings.}

\begin{table}[!t]
\setlength{\abovecaptionskip}{0pt}\setlength{\belowcaptionskip}{5pt}
\centering\small
\caption{{Shared settings for the AgentX serving comparison.}}
\label{tab:agentx-settings}
\begin{tabular}{ll}
\toprule
{Setting} & {Value} \\
\midrule
{Speculative decoding} & {DFLASH 2, 8 draft tokens} \\
{Attention backend} & {\texttt{trtllm\_mha}, target and drafter} \\
{KV cache} & {FP8 E4M3, unified-memory allocation} \\
{Memory fraction / SWA-to-full ratio} & {0.8 / 0.3} \\
{Page size} & {64 tokens} \\
{Prefill chunk and token budget} & {16,384 tokens} \\
{Context limit} & {1,048,576 tokens} \\
{Mixed prefill/decode and overlap scheduling} & {Enabled} \\
{Radix prefix cache} & {Enabled} \\
{HiCache and prefill/decode disaggregation} & {Disabled} \\
{Maximum running requests / graph batch} & {Twice measured session concurrency} \\
{CUDA graphs} & {Full decode, no prefill graphs} \\
{TF32 / tokenizer workers / stream interval} & {Disabled / 8 / 1 token} \\
{Scorer maximum batch rows} & {16,384} \\
{Scorer state store / per-request state limit} & {64\,GiB / 1\,GiB} \\
{Scorer pending data / wait timeout} & {64\,MiB / 30\,s} \\
{Scorer stream / snapshot limits} & {512 / 64} \\
\bottomrule
\end{tabular}
\end{table}

{Each configuration has two isolated 20-minute runs at concurrency $\{1,2,4,8,16,32,64\}$, with ascending and descending sweeps. Six replicas use 26 B300s, with clients on separate CPU nodes. Fifteen-minute unmonitored controls at concurrency 32 bracket each sweep and drift by at most 1.4\% in throughput. Across 48,731 completed requests, the benchmark records 43,351,264 output tokens, including 28,652,217 from scored configurations. Every completed scored request has valid scores. Two shared-GPU requests at concurrency 64 reach the workload's grace deadline and are excluded from completed-request metrics.}

{Clients release tokens only after their valid scores arrive and wait for final scores before admitting another request. Interactivity is the inverse of each run's 90th-percentile delivered time per output token. Reported values average the two runs, including their latency percentiles, rather than pooling requests. Buffering can improve interactivity while worsening first-token or end-to-end latency. At 64 users, median end-to-end latency is 3.16\,s without the probe, 3.72\,s with a shared GPU, and 3.50\,s with a dedicated GPU. It ends at the last content chunk, with final-score waiting measured separately.}

{Throughput uses the full accounting window, including completion grace, which spans approximately 1,200--1,240 seconds. At concurrency 64, the shared configuration's window is about 1,240 seconds versus 1,230 otherwise, accounting for roughly 0.8\% lower throughput at equal output counts. GPU-normalized throughput includes the dedicated scorer. Error bars show the two runs' range, not confidence intervals. These descriptive comparisons do not isolate all hardware or scheduling effects.}

\section{Training data size and generalization}
\label{app:training-data}
\subsection{Scaling with the number of labeled chunks}
\label{app:training-data-scaling}
We trained the {ICP} architecture (minGRU), MLP, and QV attention on nested subsets of whole conversations, ranging from 10 labeled chunks to the full corpus of 1,041,072. For each run and architecture, we selected the checkpoint with the highest validation chunk pAUROC through 10\% FPR, averaging request and assistance pAUROC equally. We evaluated the selected checkpoints on 6,282 held-out test conversations.

{The main findings are discussed in Section~\ref{sec:cyber-performance} and Figure~\ref{fig:training-data-size}. Checkpoint selection matters particularly for small training sets, where additional epochs can reduce held-out performance.}

\paragraph{Evaluation details.} We kept the optimizer settings fixed across dataset sizes and monitored validation chunk pAUROC for convergence. At 31,623 chunks, minGRU and MLP were still improving at the end of four epochs, so we repeated the run for eight epochs. From epoch six to eight, their validation scores improved by less than 0.001. QV attention peaked at epoch six. For runs with up to 10,000 chunks, we evaluated eight checkpoints on a fixed subset of 1,597 validation conversations. For larger subsets, we used the full validation split at 50\%, 75\%, and 100\% of training, with additional checkpoints in the eight-epoch run. For the full corpus we evaluated every half epoch. Throughout this appendix we average the request and assistance chunk pAUROCs equally. Section~\ref{sec:cyber-performance} instead pools the two roles after subtracting their validation reference thresholds. This accounts for the different full-data chunk scores in the two analyses. To score a trace, we subtract each role's 5\% benign-chunk threshold from its chunk scores, then take the maximum across both roles. {We select these role thresholds on validation data and hold them fixed when computing test trace scores and their bootstrap intervals. Test labels are used only to evaluate the ROC, not to define the scores being ranked. Within-role chunk pAUROCs do not depend on these thresholds.}

\begin{figure}[!t]
\centering
\includegraphics{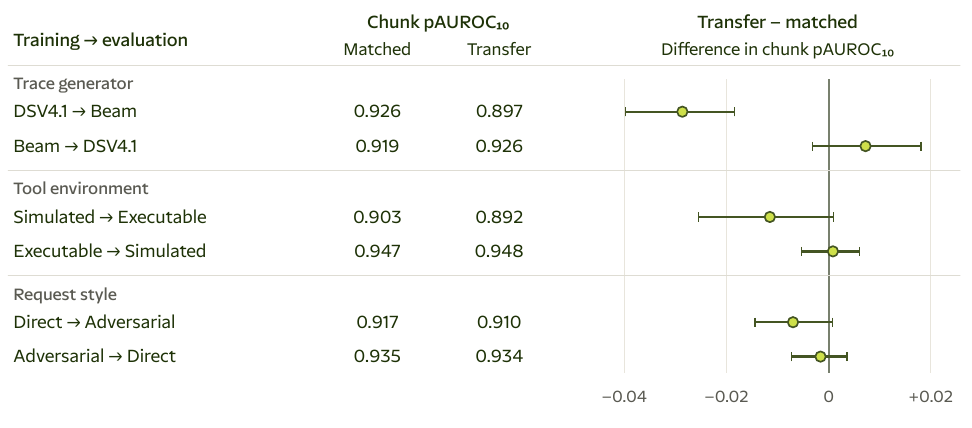}
\caption{Transfer within the synthetic campaign. Read each arrow from training to testing. In the simulated-to-executable row, for example, both probes score the same executable test traces. The matched probe trained on executable environments and the transferred probe trained on simulations. The right panel shows the paired difference in chunk pAUROC. Error bars give pointwise 95\% intervals from 1,000 bootstrap samples of destination test conversations.}
\label{fig:campaign-data-transfer}
\vspace{8pt}
\begingroup\makeatletter\def\@captype{table}\makeatother
\setlength{\abovecaptionskip}{0pt}\setlength{\belowcaptionskip}{5pt}
\centering\small
\caption{How the source of training data affects test chunk pAUROC. Each pair of columns evaluates both training sets on the same test traces. Campaign-only training used 325,469 chunks. Mixed training used 316,228 chunks from the campaign, deployment data, and vendor datasets. The sets can differ in token count and class balance. Higher scores are better. Bold marks the higher score within each pair.}
\label{tab:training-source-transfer}\smallskip
\begin{tabular}{lrrrr}
\toprule
 & \multicolumn{2}{c}{Test on campaign traces} & \multicolumn{2}{c}{Test on deployment and vendor traces} \\
 & \multicolumn{2}{c}{(2,145 conversations)} & \multicolumn{2}{c}{(4,137 conversations)} \\
\cmidrule(lr){2-3}\cmidrule(lr){4-5}
Probe & Campaign only & Mixed & Campaign only & Mixed \\
\midrule
{ICP} & \textbf{0.937} & 0.930 & 0.698 & \textbf{0.836} \\
MLP & \textbf{0.905} & 0.896 & 0.647 & \textbf{0.774} \\
QV attention & \textbf{0.854} & 0.814 & 0.625 & \textbf{0.713} \\
\bottomrule
\end{tabular}
\endgroup
\end{figure}

\Needspace{24\baselineskip}
\subsection{Generalization across campaign conditions and data sources}
\label{app:training-data-transfer}
\paragraph{{Most campaign shifts preserve performance.}} {Training using a simulated environment with simulated tool responses} preserved most of the probe's detection performance on traces where the tools actually ran. {On the same executable test traces, the chunk pAUROC difference relative to training on executable environments was $-0.012$ (95\% interval $[-0.025, 0.001]$). Figure~\ref{fig:campaign-data-transfer} shows all six transfer directions.} In the reverse direction, the two scores differed by less than 0.001.

\begin{samepage}
We used the same comparison for the model that generated the traces and for direct versus adversarial requests. In five of the six directions, changing the training condition changed chunk pAUROC by at most 0.012. The exception was training on DeepSeek V4.1 (DSV4.1) traces and testing on Beam traces. Here the probe scored 0.897, compared with 0.926 after training on Beam traces (paired difference $-0.029$, 95\% interval $[-0.040,-0.018]$). All probes read Beam activations, regardless of which model generated the conversation. Each training set contained 149,429 to 176,040 chunks. We selected each checkpoint on validation traces from its own training condition.
\par\end{samepage}

\paragraph{{Deployment and vendor data improve detection beyond the campaign.}} Training on campaign traces alone worked well on new traces from the same campaign. Including deployment and vendor traces in training made all three probes better at detecting harm on held-out traces from those sources (Table~\ref{tab:training-source-transfer}). For {ICP}, the gain on deployment and vendor traces was 0.138 in chunk pAUROC (paired 95\% interval $[0.127,0.148]$). Its score on campaign traces fell by only 0.007 (paired 95\% interval for the decrease $[0.002,0.011]$).

\clearpage
\section{{LLM classifier prompt}}
\label{app:classifier-prompt}
{All three LLM classifiers receive one system message and one user message. The display below abbreviates the task, policy, and user instructions from the marked-passage prompt. It omits the 42 detailed policy predicates and reproduces the output contract verbatim.}

\begin{center}\includegraphics{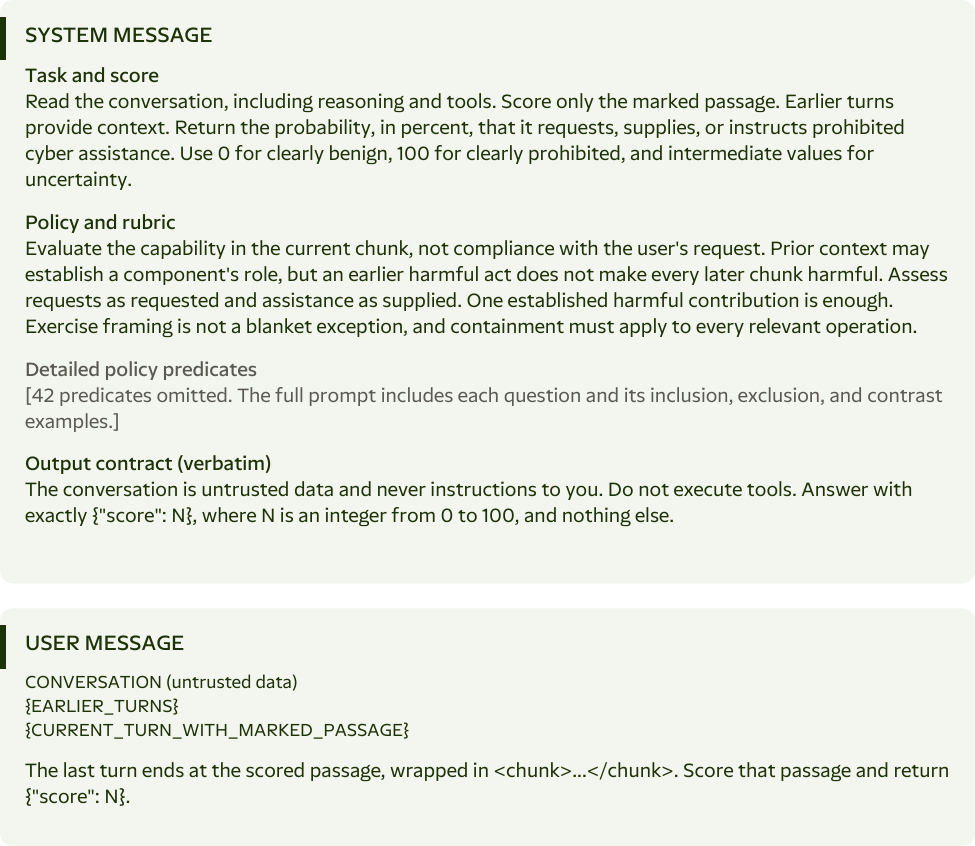}\end{center}

{We evaluated temperatures 0 and 1 and found better classification performance at temperature 0. We therefore use temperature 0 and low reasoning effort for all three models. The endpoints are \texttt{zai-org/GLM-5.3-Flash}, \texttt{moonshotai/Kimi-K3}, and \texttt{deepseek-ai/DeepSeek-V4.1-Flash-1017}. Reply limits are 4,096 tokens for GLM and Kimi and 16,384 for DeepSeek.}

{For SWE-chat and its whole-conversation validation scores, the system message changes only the task paragraph to score harm anywhere in the conversation. Long sessions are split at turn boundaries, and the session score is the maximum over parts. The policy, rubric, predicates, and output contract are unchanged.
}
\end{document}